\documentclass[preprint,11pt,3p]{elsarticle}
 
\usepackage[T1]{fontenc}
\usepackage[utf8]{inputenc}
\usepackage{amssymb}
\usepackage{amsmath}
\usepackage{latexsym}
\usepackage{mathtools}
\usepackage[format=hang,font=small,labelfont=bf]{caption}
\usepackage[outdir=./figures/]{epstopdf}
\usepackage{graphicx}
\usepackage{subfigure}
\usepackage{wasysym}
\usepackage[dvipsnames]{xcolor}
\usepackage{multirow}
\usepackage{units}
\usepackage{xcolor}
\usepackage{tikz,pgfplots}
\pgfplotsset{compat=1.17}
\usepackage[toc,page]{appendix}
\usepackage{natbib}
\usepackage{upgreek}
\usepackage{booktabs, cellspace, tabularx}
\usepackage[normalem]{ulem} 

\usepackage{multirow}
\usepackage{placeins}

\usepackage{lscape}

\newcommand{\massflux}{m}
\newcommand{\shat}{\hat{s}}
\newcommand{\ehat}{\hat{e}}
\newcommand{\chat}{\hat{c}}

\newcommand\GC{\bgroup\markoverwith
  {\textcolor{yellow}{\rule[-.5ex]{2pt}{2.5ex}}}\ULon}
\def\revone#1{{#1}}
\def\revtwo#1{{#1}}
\def\authors#1{{#1}}
\usetikzlibrary{arrows,intersections}

\graphicspath{{figures/}}

\usepackage{stmaryrd}

\newcommand{\beq}{\begin{equation}}
\newcommand{\eeq}{\end{equation}}
\newcommand{\beqs}{\begin{equation*}}
\newcommand{\eeqs}{\end{equation*}}
\newcommand{\bit}{\begin{itemize}}
\newcommand{\eit}{\end{itemize}}
\newcommand{\ben}{\begin{enumerate}}
\newcommand{\een}{\end{enumerate}}

\newcommand{\mC}{\mathcal{C}}
\newcommand{\mF}{\mathcal{F}}
\newcommand{\mR}{\mathcal{R}}

\newcommand{\mM}{\mathcal{M}}
\newcommand{\mP}{\mathcal{P}}

\newcommand{\ua}{u_{\alpha}}
\newcommand{\xa}{x_{\alpha}}

\journal{Computers and Fluids}

\begin{document}

\begin{frontmatter}

\title{Numerical treatment of the energy equation in compressible flows simulations}

\author[label1]{C. De Michele}
\author[label1]{G. Coppola}
\address[label1]{Universit\`a di Napoli ``Federico II'', Dipartimento di Ingegneria Industriale, Napoli, Italy}
\ead{gcoppola@unina.it}

\begin{abstract}
We analyze the conservation properties of various discretizations of the system of compressible Euler equations for shock-free flows, with special focus on the treatment of the energy equation and on the induced discrete equations for other thermodynamic quantities.
The analysis is conducted both theoretically and numerically and considers two important factors characterizing the various formulations, namely the choice of the energy equation and the splitting used in the discretization of the convective terms. The energy equations analyzed are total and internal energy, total enthalpy, pressure, speed of sound and entropy. In all the cases examined the discretization of the convective terms is made with locally conservative and kinetic-energy preserving schemes. Some important relations between the various formulations are highlighted and the performances of the various schemes are assessed by considering two widely used test cases. Together with some popular formulations from the literature, also new and potentially useful ones are analyzed.
\end{abstract}


\end{frontmatter}
\section{Introduction} \label{sec:Intro}

The compressible Navier-Stokes equations are written as the balance equations for mass, momentum and an `energy' variable specifying the  thermodynamic state of the system, as total or internal energy, or entropy.
The choice of the `energy' variable is usually made depending on some physical or mathematical requirement and, assumed sufficient smoothness of the flow, the various formulations are usually seen as equivalent, since one can pass from one equation to another through the usual rules of calculus and the equation of state.

It is well known that, when turning to discrete formulations, this equivalence is typically lost,
since the classical rules of calculus, which are required to pass from one set of equations to another,
cannot be applied, in general, at a discrete level~\cite{coppola2019b,Coppola_AIMETA2017}.
As an example, the product \revtwo{and chain rules} do not hold in general for \revtwo{finite-difference operators \cite{RanochaBIT2019}},
which implies that the steps required to pass from the equations for the `primary' variables
(i.e.~the balance equations directly discretized) to that for the secondary or `induced' ones
cannot be reproduced at a discrete level.
This circumstance can have strong effects on the quality of the discrete solutions,
since the derived, or induced, quantities evolve satisfying discrete equations that are, in general,
different from the discretized versions of the continuous equations.

The effects of this discrepancy are evident when considering some symmetries of the
continuous system, which are typically lost in the discrete formulation, if discretization is not properly done.
The most evident case is that of the conservation properties induced by the divergence structure of the
convective terms in the system of non-viscous equations.
In compressible flow equations the convective term
is expressed as the divergence of a flux vector.
Integration of each equation on the whole domain and application of the Gauss divergence theorem
easily shows that the convective mechanisms do not influence the evolution of
the integrated balanced quantities over the entire domain, apart from boundary terms.
The reproduction of this property at a discrete level is usually considered an important
quality of the discretization procedure.

In the case of primary variables, for which the evolution equations are directly discretized,
the divergence structure of the convective terms can be discretely enforced by
using a Finite Volume (FV) approach, which is based on the direct specification of the
flux at cell boundaries.
In this case the convective term is expressed as difference of fluxes at adjacent nodes, which is the
discrete local representation of the divergence structure.
We will term a discretization of this type a `locally conservative' discretization.
The global conservation of the quantity on the whole domain follows by virtue of the telescoping property.
In the case of a Finite Difference (FD) discretization, the divergence operator is approximated
through a suitable derivative matrix, and the local conservation form is not evident a priori.
This is especially true when an equivalent `advective' form of the convective term
(i.e.~an expression of the divergence of the product of two or more variables as
a sum of products obtained by applying the product rule) is directly discretized.
However, if one limits to the case of central schemes on uniform Cartesian meshes,
it is known that almost all the forms in which the convective terms can be written (e.g.~divergence, advective, split\dots) admit a `difference of fluxes' expression \cite{Pirozzoli2010,Fisher2013,Coppola2019}.
The extension of this and other conservation criteria to a wider class of derivative schemes, \authors{even on nonuniform meshes}, is discussed in a recent paper by Coppola and Veldman~\cite{Coppola2022}.

In the case of induced, or secondary variables, the situation is less definite.
A discretization that is locally conservative for primary variables does not
guarantee that the induced ones evolve by satisfying a discrete equation
in which the convective terms can be cast as a difference of fluxes.
As an example, the direct discretization of the system of equations for
mass, momentum and total energy
through a locally conservative formulation,
guarantees that these quantities are locally (and globally)
conserved, but the kinetic or internal energies, or the entropy, usually evolve satisfying a discrete equation
in which the convective terms cannot be cast as difference of fluxes, which means that local (and global)
preservation is spuriously affected by discrete convective terms, in a  potentially unbounded manner.

The case of kinetic energy is of particular importance, and it has been the subject of several studies
in past years, for both incompressible and compressible flows and for temporal and spatial discretizations \cite{Morinishi1998,Verstappen2003,subbareddy2009,Morinishi2010,Reiss2013,Reiss2015,Capuano2015a,Capuano2015b,capuano2017explicit,Veldman2019,Edoh2022}.
The reason for this interest lies in the fact that for incompressible
flows (global) kinetic energy is a norm of the solution vector.
A procedure that is able to bound the global kinetic energy gives also an important nonlinear stability
criterion for the discrete equations. The extension to (smooth) compressible flows
has been pursued mainly by analogy, and has shown great increases in the robustness of the simulations.
General criteria for the preservation of global kinetic energy in compressible flows equations have been recently derived for both
FD \cite{Coppola2022} and FV \cite{Veldman2019} methods.
The details of these theories will be recalled in
the subsequent Section \ref{sec:KEP}. For now, it is sufficient to mention that a globally kinetic-energy preserving (KEP) discretization involves a coordinated treatment of the convective terms in
the mass and momentum equations, without any prescription on the discretization
of the energy equation.

The situation regarding the energy equation is also interesting,
although less studied.
It is clear that a locally conservative discretization of the total-energy equation
guarantees that total energy is discretely preserved, both locally and globally,
by convection. If a KEP discretization has also been adopted
for mass and momentum equations, global preservation of internal energy follows
as a reward, but entropy usually evolves satisfying a discrete equation which is
not in locally (nor globally) conservative form.
This means that the complete discretization satisfies the first principle of thermodynamics, but
fails to satisfy the second. The reverse is true if one starts directly by discretizing
the equation for entropy. In this case a locally conservative formulation is able to exactly preserve
the entropy balance both locally and globally, but conservation of total (and internal)
energy is usually lost.
The situation in the case of a direct discretization of one of the other variables
(e.g.~internal energy, pressure, enthalpy, sound speed\dots) is even more complicated, since, in principle,
neither total energy nor entropy are preserved by convection, if one does not
properly design the discretization details.
Typically, and in absence of more suitable guidelines, the `energy' equation
(whichever one is considered among the mentioned ones)
is discretized by using a KEP formulation \revtwo{as it is done for momentum equation}, which implies the exotic global
preservation by convection of quantities such
as $\rho E^2, \rho e^2$ or $\rho s^2$,
in case the equation for total or internal energy, or entropy, respectively, is directly discretized.

In the subsequent sections we will analyze some  of the most common approaches
used in the literature in past years, together with some new formulations.
Each formulation is characterized by at least two factors.
The first is the choice of the `energy' equation to be directly discretized among the
various possibilities mentioned.
The second is the particular splitting which is used to discretize the
various equations.
It is known that both factors can strongly affect the robustness of the simulation
in different test cases, and a complete study assessing the advantages and
disadvantages of the various options has not been made yet.
In all the cases considered in this paper we will always assume that a locally
conservative and KEP discretization is performed, since these two characteristics
have been widely accepted as mandatory for a robust and reliable numerical simulation
of turbulent compressible flows.
\revone{The analysis will be mainly developed by using a classical FD formalism based on (central) discretization of divergence, advective and split forms, as in \cite{Coppola2019}.
Since in all cases these formulations can be shown to be locally conservative, explicit numerical fluxes (also of high order) will be derived.
The preference for the FD formalism stems from the fact that in this framework one can directly use the quite general necessary and sufficient condition for kinetic energy preservation developed in \cite{Coppola2019}, which is valid also in the high-order case.
However, all the numerical discretizations here analyzed can be reformulated in terms of numerical fluxes and an equivalent treatment could have been developed starting from a FV perspective.
}

It is worth mentioning that, in the context of FV methods, the theory of entropy variables \cite{Tadmor1987,Tadmor2003}
allows the specification of the conditions for Entropy Conservative \revtwo{(EC)} numerical fluxes,
which can be enforced together with the conditions for kinetic-energy preservation
to construct explicit centered numerical fluxes which are both entropy conservative
and also preserve kinetic energy for semi-discrete FV methods \revtwo{\cite{Chandrashekar2013,Ismail2009,Ranocha2020,Ranocha2021}}.
This theory, however, is based on a specification of the fluxes which typically uses a logarithmic
mean value \cite{Ismail2009}, which renders problematic the possibility of recasting
the method as a classical FD scheme
\revtwo{based on the direct discretization of divergence and advective forms as in \cite{Coppola2019}.
Moreover, the EC schemes using the logarithmic mean have some implementation issues (they need a treatment to avoid division by zero) and a non negligible increase in computational cost when compared to classical FD schemes \cite{Tamaki2022}.}
Since we \revtwo{mainly} rely on FD formulations in this work, \revtwo{which already produce many different alternatives,}  we will not consider
this approach here, \revtwo{leaving its analysis and a fairer comparison with standard FD approaches for future work.}
In Sec.~\ref{sec:KEP} we recall some of the most important ingredients of the locally
conservative and KEP discretizations, whereas in Sec.~\ref{sec:Analysis} we will analyze the
different formulations for the energy equation.
In Sec.~\ref{sec:NumRes} numerical tests on various formulations are reported for two test cases widely  used in the literature. Concluding remarks are given in Sec.~\ref{sec:Conclusions}.

\section{Problem formulation}
\label{sec:ProbForm}
\subsection{Euler equations}
The compressible Euler equations can be written as
\begin{align}
\dfrac{\partial \rho}{\partial t} &= -\dfrac{\partial \rho u_{\alpha}}{\partial x_{\alpha}} \;, \label{eq:Mass} \\[3pt]
\dfrac{\partial \rho u_{\beta}}{\partial t} &= -\dfrac{\partial \rho u_{\alpha}u_{\beta}}{\partial x_{\alpha}} -\dfrac{\partial p}{\partial x_{\beta}}  \;, \label{eq:Momentum} \\[3pt]
\dfrac{\partial \rho E}{\partial t} &=  -\dfrac{\partial \rho u_{\alpha}E}{\partial x_{\alpha}} -\dfrac{\partial p u_{\alpha}}{\partial x_{\alpha}}  \; \label{eq:TotEnergy}
\end{align}
where $\rho$ is the density, $u_{\alpha}$ is the Cartesian velocity component, $p$ is the pressure and $E$ the
total energy per unit mass, which is the sum of internal and kinetic
energies: $E = e + u_{\alpha}u_{\alpha}/2$.
The ideal gas law is assumed, which implies $p = \rho R T$ and $e = c_vT$, where $T$ is the
temperature, $R$ the gas constant and $c_v$ the specific heat at constant volume.
The ratio of specific heats at constant pressure and volume $\gamma = c_p/c_v$
is assumed to be $1.4$.
In Eq.~\eqref{eq:Mass}--\eqref{eq:TotEnergy} and in what follows we will assume the
convention that Greek subscripts refer to the components of Cartesian vectors; e.g.~$u_{\alpha}$ is the component of the velocity vector
along the $\alpha$-direction with coordinate $x_{\alpha}$ ($\alpha = 1,2,3$).
Latin subscripts as $i,j$ or $k$ are used to denote the values of the discretized
variable on a nodal point $x_i$.
When the Greek subscript is omitted (e.g.~for quantities as $u$ or $x$) it is assumed
that the relations hold for a generic value of it.
In all cases, unless otherwise explicitly stated, the summation convention over repeated Greek indices is assumed.

Equations~\eqref{eq:Mass}--\eqref{eq:TotEnergy} constitute a set of
three partial differential equations (the second one being vectorial) expressing the balance of
mass, momentum and total energy. Together with the equation of state,
they describe the evolution of both kinematic and thermodynamic
variables for an inviscid compressible flow.
In what follows, we will consider also the induced balance equations
for various quantities related to the primary variables $\rho$, $\rho u_{\alpha}$ and $\rho E$.
These equations are termed \emph{induced} because they are derived
through Eqs.~\eqref{eq:Mass}--\eqref{eq:TotEnergy}
and don't constitute additional independent balance equations.
Examples of kinematic and/or thermodynamic quantities of interest are the
kinetic energy (per unit volume) $\rho \kappa = \rho u_{\alpha}u_{\alpha}/2$,
the internal energy $\rho e$, the pressure $p$, the total enthalpy $\rho H = \rho E + p$, the sound speed $c = \sqrt{\gamma RT}$ and the entropy $\rho s=\rho c_v\ln(p/\rho^{\gamma})$.
The balance equations for these quantities are easily derived
by combining Eqs.~\eqref{eq:Mass}--\eqref{eq:TotEnergy}, together with the equation of state,
and by applying the usual rules of calculus
(assumed valid for smooth solutions), namely the classical chain and product rules of differentiation,
with respect to both temporal and space variables.
They can be written as
\begin{align}
\dfrac{\partial \rho \kappa}{\partial t} &= -\dfrac{\partial \rho u_{\alpha} \kappa}{\partial x_{\alpha}} -u_{\alpha}\dfrac{\partial p}{\partial x_{\alpha}}\;, \label{eq:KinEnergy} \\[3pt]
\dfrac{\partial \rho e}{\partial t} &= -\dfrac{\partial \rho u_{\alpha} e}{\partial x_{\alpha}} -p\dfrac{\partial u_{\alpha}}{\partial x_{\alpha}}  \;, \label{eq:IntEnergy} \\[3pt]
\dfrac{\partial p}{\partial t} &=  -\dfrac{\partial p u_{\alpha}}{\partial x_{\alpha}} -\left(\gamma -1\right)p\dfrac{\partial u_{\alpha}}{\partial x_{\alpha}}  \;, \label{eq:Pressure}\\[3pt]
\dfrac{\partial \rho H}{\partial t} &= -\dfrac{\partial \rho u_{\alpha} H}{\partial x_{\alpha}} -
\dfrac{\partial p u_{\alpha}}{\partial x_{\alpha}} -\left(\gamma -1\right)p\dfrac{\partial u_{\alpha}}{\partial x_{\alpha}}  \;, \label{eq:TotEnthalpy}\\[3pt]
\dfrac{\partial \rho c}{\partial t} &= -\dfrac{\partial \rho u_{\alpha} c}{\partial x_{\alpha}} -\frac{\left(\gamma -1\right)}{2}\rho c\dfrac{\partial u_{\alpha}}{\partial x_{\alpha}}  \;, \label{eq:SoundSpeed}\\[3pt]
\dfrac{\partial \rho s}{\partial t} &= -\dfrac{\partial \rho u_{\alpha} s}{\partial x_{\alpha}} \;. \label{eq:Entropy}
\end{align}

On a continuous ground, Eqs.~\eqref{eq:KinEnergy}--\eqref{eq:Entropy}
are always satisfied by the variables obtained as a combination of
the solutions to Eqs.~\eqref{eq:Mass}--\eqref{eq:TotEnergy},
once a sufficient smoothness has been assumed.
In principle, any of the  Eqs.~\eqref{eq:IntEnergy}--\eqref{eq:Entropy}
can be used in place of Eq.~\eqref{eq:TotEnergy}, to describe the evolution
of the system (note  that Eq.~\eqref{eq:KinEnergy},
being obtained by combining only
Eqs.~\eqref{eq:Mass} and \eqref{eq:Momentum}, is independent of the equation
for total energy, and cannot be used in place of it).
To each choice of the `energy' equation corresponds a set of `primary' variables,
and the values of the other `induced' ones can  be obtained by algebraic
manipulations and through the equation of state.

\subsection{Discrete approximations}\label{sec:Discrete}
In this paper we will assume that the equations of motion are
discretized with a FD 
method
over a uniform Cartesian mesh of width $h$ (with a colocated approach).
We will also assume that integration is performed
through a semi-discretized
approach, in which a spatial discretization step is firstly performed, and the
resulting system of Ordinary Differential Equations (ODE) is integrated in time by
using a standard solver.
Since we focus on the space discretization step, we will assume that all the
manipulations involving time derivatives can be carried out at the continuous level.
The effects of time integration errors will be assumed to be negligible
at sufficiently small time steps.
Spatial discretization is made by using central difference schemes which,
among various important properties, assure that the discrete counterpart of the
integration by parts rule (i.e.~the summation by parts (SBP) rule)
holds, for periodic boundary conditions \cite{Mansour1979}.
Of course, SBP operators can be derived also for non-periodic
boundary conditions. In this case, all the reasonings which are based on
the SBP rule hold in the general case.
In the derivation of the various properties of the discrete equations,
manipulation of spatial terms will be done by using only algebraic relations and the SBP rule,
whereas the product and chain rules of derivative will not be allowed, since they are not valid, in general, for discrete operators.
Under these assumptions,
all the equations derived from the primary ones will be valid at discrete level.

To distinguish between continuous and discrete operators, we use the symbol $\delta$ for discrete derivatives,
in contrast to the usual symbol $\partial$ for partial derivatives.
According to the previous discussion, for discrete operators we will assume all the usual algebraic operations
valid for derivative operators, including the SBP rule, but the product rule will not be allowed.
The result of manipulations with $\delta$ operators will hold also on a continuous ground, but the opposite, of course, is not true.
To be concrete, the $\delta/\delta x$ operator is typically a central (explicit)
derivative scheme on uniform mesh, of the form
$\delta\phi_i/\delta x = \sum_{k = 1}^L a_k\left(\phi_{i+k}-\phi_{i-k}\right)/h$, for which the
classical product rule
$\delta\rho\phi/\delta x =\rho\delta\phi/\delta x +\phi\delta\rho/\delta x$ does not hold.
However, for such operators the SBP rule with periodic boundary conditions
$$\sum_i \rho_i\frac{\delta\phi_i}{\delta x}h = -\sum_i \phi_i\frac{\delta\rho_i}{\delta x}h$$
is easily shown to hold~\cite{coppola2019b,Coppola_AIMETA2017,Kravchenko1997}.

Discrete convective terms will be analyzed in their property to be globally or locally conservative.
A globally conservative discretization is such that the sum over the grid points of the discretized formula is zero for periodic or homogeneous boundary conditions.
Locally conservative discretization, on the other hand, are such that each individual discretization can be expressed as difference of fluxes at adjacent nodes.
Of course, local conservation implies global conservation.
The opposite implication is also true (although less trivial) for a wide class of approximations of the convective terms appearing in Eq.~\eqref{eq:Mass}-\eqref{eq:Entropy}, as it is shown in the recent paper by Coppola and Veldman~\cite{Coppola2022}.

\section{Kinetic-energy preserving formulations} \label{sec:KEP}
\subsection{Discrete evolution of the generalized kinetic energy }\label{sec:Discrete_KEP_evol}
Equations.~\eqref{eq:Mass}--\eqref{eq:Entropy}
have the general structure:
\begin{equation}\label{eq:GenStructBalEq}
    \dfrac{\partial \rho\phi}{\partial t}=-\mathcal{R}_{\rho\phi}=-\mathcal{C}_{\rho\phi}-\mathcal{P}_{\rho\phi}
\end{equation}
where $\mathcal{C}_{\rho\phi}$ is the convective term, in the form
of the divergence of a convective flux,
and $\mathcal{P}_{\rho\phi}$ is a pressure term.
The symbols $\mR, \mC, \mP$ will be used here to denote both the individual spatial
terms at the right hand sides of Eqs.~\eqref{eq:Mass}-\eqref{eq:Entropy},
or their spatial discretizations, as in the
Eq.~\eqref{eq:Mass_Forms}, \eqref{eq:Mom_Forms} and \eqref{eq:ConvKinGlobPres} below,
the correct interpretation emerging from the context.
Under the assumptions mentioned in Sec.~\ref{sec:Discrete}, by manipulating Eq.~\eqref{eq:Mass} and \eqref{eq:GenStructBalEq}
the induced discrete evolution equation
for the generalized kinetic energy $\rho\phi^2/2$ can be written as~\cite{coppola2019b,Coppola2019,Veldman2019} (see also Sec.~\ref{sec:GenForm})
\begin{equation}
    \dfrac{\partial\rho \phi^2/2}{\partial t} = - \left(\phi\mathcal{C}_{\rho\phi}-\dfrac{\phi^2}{2}\mathcal{M}\right) -
    \phi\mathcal{P}_{\rho\phi},\label{eq:GenKinEnergy_a}
\end{equation}
where $\mM$ is a special symbol we use to denote $\mC_{\rho}$.
The case $\phi = u_{\alpha}$ gives the induced equation for the classical kinetic energy
per unit volume $\rho u_{\alpha}^2/2$.

The condition that the generalized kinetic energy is preserved (locally or globally) by
convective terms amounts to the requirement
that the term $\mathcal{C}_{\rho \phi^2/2} = \phi\mathcal{C}_{\rho\phi}-(\phi^2/2)\mathcal{M}$
is in (local or global) conservation form.
\revtwo{We explicitly note that our definition of (generalized) kinetic-energy
preserving discretization refers to the convective term in the discrete equation
for $\rho\phi^2/2$, which puts requirements only on $\mC_{\rho\phi}$ and $\mM$.
The discretization of the pressure term in the equation for $\rho\phi$ (which influences the term $\phi\mP_{\rho\phi}$ in Eq.~\eqref{eq:GenKinEnergy_a}) is left outside
the definition of KEP discretization.
This could introduce some indeterminacy in the notion of KEP discretization, as any consistent convection term $\mC_{\rho\phi}$ could be split into a KEP contribution plus some additional term to be included into the pressure term without altering its consistency (for a discussion on a related topic see \cite{Ranocha2018}).
However, we implicitly assume that the pressure term is a straightforward discretization of its continuous counterpart. In all the applications we present in this work the pressure term in the momentum equation is a simple central discretization of the gradient of $p$.}

The analysis reported in \cite{Coppola2019}
shows that by using suitable discretizations for the terms $\mathcal{M}$ and $\mathcal{C}_{\rho\phi}$, the convective term $\mC_{\rho\phi^2/2}$ comes
automatically in conservation form. In fact,
a one-parameter family of locally-conservative and kinetic-energy preserving
forms is possible.

This family can be easily written by defining the divergence and advective forms
for the convective term in the mass equation as:
\begin{equation}\label{eq:Mass_Forms}
    \mathcal{M}^D = \dfrac{\delta \rho u_{\alpha}}{\delta x_{\alpha}},\qquad
    \mathcal{M}^A = \rho\dfrac{\delta u_{\alpha}}{\delta x_{\alpha}}+u_{\alpha}\dfrac{\delta \rho}{\delta x_{\alpha}}
\end{equation}
and by using the following expressions for the convective terms in the momentum equation~\cite{Coppola2019}:
\begin{equation}\label{eq:Mom_Forms}
    \mathcal{C}_{\rho\phi}^D = \dfrac{\delta \rho u_{\alpha} \phi}{\delta x_{\alpha}}, \quad
    \mathcal{C}_{\rho\phi}^{\phi} = \phi \dfrac{\delta \rho u_{\alpha}}{\delta x_{\alpha}} + \rho u_{\alpha}\dfrac{\delta \phi}{\delta x_{\alpha}},\quad
    \mathcal{C}_{\rho\phi}^{u} = u_{\alpha}\dfrac{\delta \rho \phi}{\delta x_{\alpha}}+\rho \phi\dfrac{\delta u_{\alpha}}{\delta x_{\alpha}},\quad
    \mathcal{C}_{\rho\phi}^{\rho}=\rho \dfrac{\delta u_{\alpha} \phi}{\delta x_{\alpha}} + \phi u_{\alpha}\dfrac{\delta \rho}{\delta x_{\alpha}}.
\end{equation}
By combining these expressions, the
Feiereisen \emph{et al.} \cite{Feiereisen1981} and
Coppola \emph{et al.} \cite{Coppola2019} forms are defined as
\begin{equation}\label{eq:F-C_Forms}
\mC_{\rho\phi}^F = \dfrac{\mC_{\rho\phi}^{D}+\mathcal{C}_{\rho\phi}^{\phi}}{2},\;\qquad
\mC_{\rho\phi}^C=\dfrac{\mC_{\rho\phi}^{u}+\mathcal{C}_{\rho\phi}^{\rho}}{2}\;.
\end{equation}
With these definitions one can easily show that the one-parameter family of forms
\begin{align}
\mM &= \xi\mM^D + (1-\xi)\mM^A\;,\label{eq:Mdecomp}\\
\mC_{\rho\phi} &= \xi\mC_{\rho\phi}^F + (1-\xi)\mC_{\rho\phi}^C\;\label{eq:Cdecomp}
\end{align}
is kinetic-energy preserving.
This means that, whatever the value of $\xi$ is, the discrete terms $\mM$
and $\mC_{\rho\phi}$ defined by Eqs.~\eqref{eq:Mdecomp} and \eqref{eq:Cdecomp}
induce a conservative structure for the term
$\mathcal{C}_{\rho \phi^2/2}$
when central schemes on uniform meshes are used.
This result has been recently extended to arbitrary (i.e.~non central) schemes
on  non-uniform (Cartesian) meshes in \cite{Coppola2022}.

The cases $\xi=0$ and $\xi=1$ give the classical Feiereisen form and the newly
derived form described in \cite{Coppola2019}. The case $\xi = 1/2$ furnishes
the form investigated by Kennedy and Gruber~\cite{Kennedy2008},
which was shown to be energy preserving by Pirozzoli \cite{Pirozzoli2010}
and which is here denoted as the Kennedy-Gruber-Pirozzoli (KGP) form.

\subsection{Global conservation}
Global conservation is easily shown by substituting
Eqs.~\eqref{eq:Mass_Forms} and \eqref{eq:Mom_Forms}
into Eq.~\eqref{eq:F-C_Forms} and Eqs.~\eqref{eq:Mdecomp}-\eqref{eq:Cdecomp}.
The convective term for the kinetic energy $\mathcal{C}_{\rho\phi^2/2}$ in Eq.~\eqref{eq:GenKinEnergy_a} eventually reads
\begin{equation}\label{eq:ConvKinGlobPres}
    \mC_{\rho\phi^2/2} = \dfrac{\xi}{2}\left(\phi\dfrac{\delta \rho\phi u_{\alpha}}{\delta x_{\alpha}} + \rho\phi u_{\alpha}\dfrac{\delta \phi}{\delta x_{\alpha}}\right) +
\dfrac{1-\xi}{2}\left(u_{\alpha} \phi\dfrac{\delta \rho  \phi}{\delta x_{\alpha}} + \rho\phi\dfrac{\delta u_{\alpha}\phi}{\delta x_{\alpha}}\right).
\end{equation}
Summation over the entire domain and application of the SBP
property easily shows that for homogeneous or periodic boundary conditions the
sums within the parentheses individually vanish, showing global conservation
independently of the value of $\xi$;
the validity of the SBP rule is crucial in showing global preservation of generalized kinetic energy.

Note that the right hand side of Eq.~\eqref{eq:ConvKinGlobPres} can be expressed in matrix-vector notation by defining the grid vectors $\uprho,\mathsf{u}$ and $\upphi$, gathering the individual mesh values $\rho_i,u_i$ and $\phi_i$,
and the global derivative matrix $\mathsf{D}$ containing the weights of the derivative formula $a_k$.
With this notation the term $\mC_{\rho\phi^2/2}$ is a vector expressed by
\beq\label{eq:ConvKinGlobPres_Mat1}
\dfrac{\xi}{2}\left(\Phi\mathsf{D}\mathsf{R}\mathsf{U}\upphi +
\Phi\mathsf{U}\mathsf{R}\mathsf{D}\upphi\right) +
\dfrac{1-\xi}{2}\left(\Phi\mathsf{U}\mathsf{D}\mathsf{R}\upphi +
\Phi\mathsf{R}\mathsf{D}\mathsf{U}\upphi\right).
\eeq
where $\Phi = \text{diag}(\upphi)$, $\mathsf{U} = \text{diag}(\mathsf{u})$ and
$\mathsf{R} = \text{diag}(\uprho)$.
Integration in space over the uniform mesh
is equivalent, in this notation, to the sum of the components of the vector,
expressed through the quadratic form
\beq\label{eq:ConvKinGlobPres_Mat2}
\upphi^T\left[\dfrac{\xi}{2}\left(\mathsf{D}\mathsf{R}\mathsf{U} +
\mathsf{U}\mathsf{R}\mathsf{D}\right) +
\dfrac{1-\xi}{2}\left(\mathsf{U}\mathsf{D}\mathsf{R} +
\mathsf{R}\mathsf{D}\mathsf{U}\right)\right]\upphi.
\eeq
Skew symmetry of the matrix within square brackets is a necessary and sufficient condition
for global conservation of generalized kinetic energy
\cite{Morinishi2010,Coppola2019,coppola2019b,Veldman2019} and is equivalent to the Requirement 3.1
mentioned in the recent paper by Veldman~\cite{Veldman2021}.
This property is guaranteed by the skew-symmetry of the derivative matrix $\mathsf{D}$,
which is the matrix-vector version of the SBP property in our context.

It is interesting to note that, although a variable
coefficient $\xi(\mathbf{x})$ could have been admitted for the definition of
consistent and locally-conservative approximations of mass and
momentum convective terms,
in Eq.~\eqref{eq:Mdecomp} and \eqref{eq:Cdecomp}
a constant value of $\xi$ is assumed on the whole domain,
since a variable coefficient $\xi(\mathbf{x})$ would have invalidated
the proof of global preservation
of generalized kinetic energy based on Eq.~\eqref{eq:ConvKinGlobPres} and
\eqref{eq:ConvKinGlobPres_Mat1}--\eqref{eq:ConvKinGlobPres_Mat2}.
In what follows we will see that a locally conservative approximation
depending on a variable coefficient, which also preserves
the generalized kinetic energy, can be constructed in the framework of
a FV formulation of the terms $\mM$ and $\mC_{\rho \phi}$.

\subsection{Local conservation}\label{sec:LocalConservation}
To show local conservation we can here use the fact that all the forms in
Eqs.~\eqref{eq:Mass_Forms}--\eqref{eq:F-C_Forms}
have a locally conservative expression when central schemes are used,
i.e.~at a generic node $x_i$ they can
be written as the difference of numerical fluxes $(\hat{F}_{i+1/2}-\hat{F}_{i+1/2})/h$.
As shown in \cite{Pirozzoli2010} and \cite{Coppola2019}, the numerical
flux $\hat{F}_{i+1/2}$ has the general form
\beq
\hat{F}_{i+1/2} = 2\sum_{k=1}^La_k\sum_{m=0}^{k-1}\mathcal{I}\left(\rho,u,\phi\right)_{i-m,k}, \label{eq:Flux3}
\eeq
where $a_k$ are the coefficients of a central and explicit differentiation formula
and $\mathcal{I}\left(\rho,u,\phi \right)_{i,k}$ is a suitable interpolation operator.
The list of the interpolation operators associated with the forms~\eqref{eq:Mass_Forms}-\eqref{eq:Mom_Forms} is~\cite{Coppola2019}:
\begin{equation}\label{eq:InterpOp}
\begin{array}{lclclclc}
\mM^D            & \longrightarrow & \mathcal{I}(\rho,u)_{i,k}     &=& \overline{\rho u}^\revtwo{{i+k/2}}                      &=& \dfrac{(\rho u)_{i}+(\rho u)_{i+k}}{2},\\[4pt]
\mM^A            & \longrightarrow & \mathcal{I}(\rho,u)_{i,k}     &=& \overline{\overline{\left(\rho, u\right)}}^\revtwo{{i+k/2}}           &=& \dfrac{\rho_i u_{i+k}+\rho_{i+k} u_{i}}{2},\\ [4pt]
\mC_{\rho\phi}^{D} & \longrightarrow & \mathcal{I}(\rho,u,\phi)_{i,k}&=& \overline{\rho u\phi}^\revtwo{{i+k/2}} &=& \dfrac{(\rho u\phi)_{i}+(\rho u\phi)_{i+k}}{2},\\[4pt]
\mC_{\rho\phi}^{\phi} & \longrightarrow & \mathcal{I}(\rho,u,\phi)_{i,k}&=& \overline{\overline{\left(\rho u,\phi\right)}}^\revtwo{{i+k/2}} &=& \dfrac{\left(\rho u\right)_{i}\phi_{i+k}+\left(\rho u\right)_{i+k}\phi_i}{2},\\[4pt]
\mC_{\rho\phi}^{u} & \longrightarrow & \mathcal{I}(\rho,u,\phi)_{i,k}&=& \overline{\overline{\left(\rho \phi,u\right)}}^\revtwo{{i+k/2}} &=& \dfrac{\left(\rho \phi\right)_{i}u_{i+k}+\left(\rho \phi\right)_{i+k}u_i}{2},\\[4pt]
\mC_{\rho\phi}^{\rho} & \longrightarrow & \mathcal{I}(\rho,u,\phi)_{i,k}&=& \overline{\overline{\left(\phi u,\rho\right)}}^\revtwo{{i+k/2}} &=& \dfrac{\left(\phi u\right)_{i}\rho_{i+k}+\left(\phi u\right)_{i+k}\rho_i}{2},
\end{array}
\end{equation}
from these relations the interpolations associated to $\mC^F, \mC^C$ and $\mC^{KGP}$ are easily obtained:
\begin{equation}\label{eq:InterpOp_Cforms}
\begin{array}{lclclclc}
\mC_{\rho\phi}^F & \longrightarrow & \mathcal{I}(\rho,u,\phi)_{i,k}&=& \overline{\phi}^\revtwo{{i+k/2}}\overline{\rho u}^\revtwo{{i+k/2}} &=& \left(\dfrac{\phi_{i}+\phi_{i+k}}{2}\right)\dfrac{(\rho u)_{i}+(\rho u)_{i+k}}{2},\\[4pt]
\mC_{\rho\phi}^C & \longrightarrow & \mathcal{I}(\rho,u,\phi)_{i,k}&=& \overline{\phi}^\revtwo{{i+k/2}}\overline{\overline{\left(\rho, u\right)}}^\revtwo{{i+k/2}} &=&\left(\dfrac{\phi_{i}+\phi_{i+k}}{2}\right)\dfrac{\rho_i u_{i+k}+\rho_{i+k} u_{i}}{2},\\[4pt]
\mC_{\rho\phi}^{KGP} & \longrightarrow & \mathcal{I}(\rho,u,\phi)_{i,k}&=& \overline{\phi}^\revtwo{{i+k/2}}\overline{\rho}^\revtwo{{i+k/2}}\overline{u}^\revtwo{{i+k/2}} &=&\left(\dfrac{\phi_{i}+\phi_{i+k}}{2}\right)\left(\dfrac{\rho_{i}+\rho_{i+k}}{2}\right)\left(\dfrac{u_{i}+u_{i+k}}{2}\right).
\end{array}
\end{equation}
For the sake of clarity, but without loss of generality, from now on we work on
the simpler second-order case, for which
$L=1$, $a_1=1/2$, $\overline{\phi}^\revtwo{{i+1/2}} =  (\phi_{i+1}+\phi_i)/2$
and $\overline{\overline{\left(\rho, u\right)}}^\revtwo{{i+1/2}} = (\rho_i u_{i+1}+\rho_{i+1}u_{i})/2$.
To further simplify notation we will drop the superscript \revtwo{{$i+1/2$}} from the definition of
second-order interpolation operators, when no ambiguity can occur (\revtwo{i.e. $\overline{\phi}^{i+1/2} = \overline{\phi}\,$}).
The extension of the results here derived to the higher-order case is reported in~\ref{Sec:Appendix}.

In the second-order case, each individual interpolation operator $\mathcal{I}(\rho,u,\phi)_{i,1}$
coincides with the numerical flux $\hat{F}_{i+1/2}$ and Eqs.~\eqref{eq:Mdecomp} and \eqref{eq:Cdecomp} can be written as
\begin{align}
\mM &= \xi\llbracket\overline{\rho u}\rrbracket+(1-\xi)\llbracket\overline{\overline{\left(\rho, u\right)}}\rrbracket =
\llbracket m_{i+1/2}\rrbracket\;,\label{eq:Mdecomp2}\\
\mC_{\rho\phi} &= \xi\llbracket\overline{\phi}\,\overline{\rho u}\rrbracket+(1-\xi)\llbracket\overline{\phi}\,\overline{\overline{\left(\rho, u\right)}}\rrbracket
=\llbracket\overline{\phi}\,m_{i+1/2}\rrbracket\;\label{eq:Cdecomp2}
\end{align}
where $\llbracket\cdot\rrbracket$ is the difference operator: $\llbracket m_{i+1/2}\rrbracket = \left(m_{i+1/2}-m_{i-1/2}\right)/h$
and the mass flux $m_{i+1/2}$ is given by
\beq\label{eq:MassFlux}
m_{i+1/2} = \xi\,\overline{\rho u} + (1-\xi)\,\overline{\overline{\left(\rho, u\right)}}.
\eeq

Equations~\eqref{eq:Mdecomp2}-\eqref{eq:Cdecomp2} can be seen as the
FV formulation of the convective terms $\mM$ and $\mC_{\rho\phi}$, whereas $m_{i+1/2}$ in Eq.~\eqref{eq:MassFlux}
is the most general one-parameter symmetric bilinear approximation of the mass flux $\rho u$ over a
two-point stencil $\left\{x_i,x_{i+1}\right\}$ (cfr.~\cite{Coppola2022}).
Note that the approximation of the convective term $\mC_{\rho\phi}$
is obtained through the flux $\hat{F}_{i+1/2}=\overline{\phi} m_{i+1/2}$.
This special form of the flux is a direct consequence of the fact that we started
from a KEP approximation of $\mC_{\rho\phi}$ and is consistent with the necessary and sufficient condition for second-order KEP fluxes~\cite{Jameson2008b,Veldman2019}.
The symmetric interpolation $\overline{\phi}$ appearing in Eq.~\eqref{eq:Cdecomp2}
is not strictly associated to the uniform mesh, but survives also in KEP approximations on arbitrary (even non Cartesian) meshes.

By combining Eq.~\eqref{eq:Mdecomp2}-\eqref{eq:Cdecomp2}, the discrete convective term for the generalized kinetic energy can now be written as:
\begin{multline}
\mC_{\rho \phi^2/2} = \left(\phi_i\mathcal{C}_{\rho\phi}-\dfrac{\phi_i^2}{2}\mathcal{M}\right)=
\phi_i\left\llbracket\overline{\phi}^\revtwo{i+1/2} m_{i+1/2}\right\rrbracket-\frac{\phi^2_i}{2}\left\llbracket m_{i+1/2}\right\rrbracket\\
=\frac{1}{h}\left(\phi_i\overline{\phi}^\revtwo{i+1/2} m_{i+1/2}-\phi_i\overline{\phi}^\revtwo{i-1/2}m_{i-1/2}-
\frac{\phi_i^2}{2}\left(m_{i+1/2}-m_{i-1/2}\right)\right)=\\
\frac{1}{h}\left(\frac{\phi_i\phi_{i+1}}{2} m_{i+1/2}-
\frac{\phi_{i-1}\phi_i}{2} m_{i-1/2}\right)=
\left\llbracket\frac{\phi_i\phi_{i+1}}{2} m_{i+1/2}\right\rrbracket
\end{multline}
which shows that the convective term of the discrete equation for the generalized kinetic energy
can be written in locally conservative form with local flux
\beq\label{eq:KinEn_Flux}
\mF_{\rho\phi^2/2}=\frac{1}{2}\phi_i\phi_{i+1}m_{i+1/2}.
\eeq
This result is a particular case of the equivalence between global and local conservative formulations which has been derived in a more general framework in \cite{Coppola2022}.

\subsection{Discrete kinetic-energy evolution equation}\label{sec:Discrete_KE_Eq}
Since the above derivation shows that the generalized kinetic energy discretely
evolves according to a locally conservative formulation of the convective term,
one is left with the question of which finite-difference formulation is associated to the fluxes~\eqref{eq:KinEn_Flux}.
To investigate this aspect, we rewrite the flux $\mF_{\rho\phi^2/2}$ by using
Eq.~\eqref{eq:MassFlux} which gives
\begin{multline}
\mF_{\rho \phi^2/2} = \frac{\phi_i\phi_{i+1}}{2}\left(\xi\,\overline{\rho u} + (1-\xi)\overline{\overline{\left(\rho, u\right)}}\right) =
\frac{\xi}{2}\left(\phi_i\phi_{i+1}\,\overline{\rho u}\right) +
\frac{1-\xi}{2}\left(\phi_i\phi_{i+1}\,\overline{\overline{\left(\rho,u\right)}}\right) \\
=\frac{\xi}{2}\left(\frac{\left(\rho u\phi\right)_i\phi_{i+1} + \left(\rho u\phi\right)_{i+1}\phi_{i}}{2}\right) +
\frac{1-\xi}{2}\left(\frac{\left(\rho\phi\right)_i\left(u\phi\right)_{i+1} + \left(\rho\phi\right)_{i+1}\left(u\phi\right)_{i}}{2}\right)  \\
= \frac{\xi}{2}\,\overline{\overline{\left(\rho u \phi,\phi\right)}} +
\frac{1-\xi}{2}\,\overline{\overline{\left(\rho \phi,u\phi\right)}}.
\end{multline}
This manipulation shows, from a different perspective, the obvious result that the
flux \eqref{eq:KinEn_Flux} corresponds to a finite-difference discretization of
the convective term in the generalized kinetic-energy equation which is
built according to Eq.~\eqref{eq:ConvKinGlobPres}, which is one of the possible splittings of the derivative of the quadruple product $\rho u\phi\phi$.
In principle, one could directly discretize the equation for the generalized kinetic
energy in place of the equation for $\rho\phi$.
If the discretization of the convective term in the equation for
$\rho\phi^2/2$ is made according to Eq.~\eqref{eq:ConvKinGlobPres},
the discrete equation for $\rho\phi$ (which is now an induced equation)
has a convective term which is equivalent to Eq.~\eqref{eq:Cdecomp2}.
The two formulations (i.e.~the direct discretization of $\rho\phi$ through
Eq.~\eqref{eq:Cdecomp2}, with the consequent induced discrete evolution
of $\rho\phi^2/2$ according to Eq.~\eqref{eq:ConvKinGlobPres}, or the
direct discretization of $\rho\phi^2/2$ according to Eq.~\eqref{eq:ConvKinGlobPres})
are equivalent under the assumption of exact time integration.

This derivation can be used also as a general guideline to design a
discretization procedure for the convective term of an arbitrary
(non negative) quantity $\psi$ which induces a locally conservative discrete evolution for
a quantity $\phi \propto \sqrt{\psi}$.
We will see an application of this procedure in Sec.~\ref{sec:TotEnergy}.

As a final remark, we note that, since in the FV formulation
\eqref{eq:Mdecomp2}--\eqref{eq:MassFlux} the parameter $\xi$ is inside the mass flux
$m_{i+1/2}$, the construction of the terms $\mM$ and $\mC_{\rho\phi}$
can be now conducted by allowing a pointwise specification of the
parameter $\xi$, without spoiling the local conservation of kinetic energy.
In fact, the derivation made above can be rephrased by directly assuming
a local `weight' $\xi_i$ inside the definition of the mass flux:
$m_{i+1/2} = \xi_i\,\overline{\rho u} + (1-\xi_i)\overline{\overline{\left(\rho, u\right)}}$
and by defining $\mM$ and $\mC_{\rho\phi}$ as usual as $\mM = \left\llbracket m_{i+1/2}\right\rrbracket$
and $\mC_{\rho\phi}=\left\llbracket\,\overline{\phi}\,m_{i+1/2}\right\rrbracket$.
This specification does not affect the local conservation of linear invariants
and of the generalized kinetic energy, adding a great number of degrees of freedom
which can be optimized, also in an adaptive way, to achieve different targets.

\section{Analysis of induced discrete equations} \label{sec:Analysis}
Having established the conditions for KEP formulations,
we now move on to the analysis of the induced discrete equations associated with a
certain choice for the `energy' variable $\rho\phi$.
We will always assume that all the convective terms of the various equations are
discretized by using locally conservative formulations.
Moreover, we will assume that a KEP formulation from
Eqs.~\eqref{eq:Mdecomp}-\eqref{eq:Cdecomp}
has been used for mass and  momentum equations.

\subsection{General framework} \label{sec:GenForm}
To better investigate the relations among the various discrete formulations, we
introduce a function $G(\rho,u_{\alpha},\phi)$ representing an `induced' variable whose
balance is obtained by combining the equations for the
`primary' variables $\rho,\rho u_{\alpha}$ and $\rho\phi$.
To derive the discrete evolution equation for $G$, we assume exact temporal integration and
use the  chain rule for temporal derivatives
\begin{equation}
    \dfrac{\partial G(\rho,u_{\alpha},\phi)}{\partial t}=G_{\rho}\dfrac{\partial \rho}{\partial t}+G_{u_{\alpha}}\dfrac{\partial u_{\alpha}}{\partial t}+G_{\phi}\dfrac{\partial \phi}{\partial t},
\end{equation}
which can be written in terms of the right hand sides of the equations for
$\rho,\rho u_{\alpha}$ and $\rho\phi$ and finally furnishes:
\begin{equation}\label{eq:Gevol}
    \dfrac{\partial G}{\partial t}=-\left(G_{\rho}-\frac{u_{\alpha}}{\rho}G_{u_{\alpha}}-\frac{\phi}{\rho}G_{\phi}\right)\mathcal{M}-
    \frac{G_{u_{\alpha}}}{\rho}\mathcal{R}_{\rho u_{\alpha}} - \frac{G_{\phi}}{\rho}\mathcal{R}_{\rho\phi}.
\end{equation}
Since this equation has been obtained by using algebraic relations and by
manipulating only temporal derivatives, it holds also at a discrete level
(for exact time integration)
and gives the induced discrete evolution equation for any quantity $G$
as a function of the discrete spatial terms of the balance equations for the
primary quantities $\rho,\rho u_{\alpha}$ and $\rho\phi$.

As a simple example of the  application of this formula, we consider again the case
of the kinetic energy, which is a function of $\rho$ and $u_{\alpha}$:
$G(\rho,u_{\alpha}) = \rho u_{\alpha}^2/2$. In this case one has $G_{\rho} = u_{\alpha}^2/2$,
$G_{u_{\alpha}} = \rho u_{\alpha}$ and $G_{\phi}=0$ and application of Eq.~\eqref{eq:Gevol} directly gives
\begin{equation}
    \dfrac{\partial\rho u_{\alpha}^2/2}{\partial t} =
    - \left(u_{\alpha}\mathcal{R}_{\rho u_{\alpha}}-\dfrac{u_{\alpha}^2}{2}\mathcal{M}\right)\label{eq:KinEnergy_a}
\end{equation}
which is Eq.~\eqref{eq:GenKinEnergy_a} in the case $\phi = u_{\alpha}$.
Note that Eq.~\eqref{eq:KinEnergy_a} is valid in both the cases in which
a summation convention over the repeated index $\alpha$ is assumed or not,
which means that in a KEP formulation from Eq.~\eqref{eq:Mdecomp}--\eqref{eq:Cdecomp}
the kinetic energy is preserved by convection separately
for each contribution $\rho u_{\alpha}^2/2$.

Application of Eq.~\eqref{eq:Gevol} to arbitrary quantities $G$ as functions of
the various `energy' variables $\phi$ is detailed in the next subsections.

\subsection{Total-energy equation} \label{sec:TotEnergy}
The most commonly adopted  choice for the energy equation is
the Eq.~\eqref{eq:TotEnergy} for total energy $\rho E$.
It has been widely used in the past for compressible simulations and various formulations have been analyzed in the literature.
Among the various contributions, we mention here Jameson~\cite{Jameson2008b}, Pirozzoli~\cite{Pirozzoli2010},
Subbareddy and Candler~\cite{subbareddy2009}, Kennedy and Gruber~\cite{Kennedy2008} and Kuya \emph{et al.}~\cite{Kuya2018}.
The divergence structure of both convective
and pressure terms easily shows that total energy is globally conserved
(for homogeneous or periodic boundary conditions)
and that the local variation inside a cell is driven only by boundary terms.
This property can be reproduced at discrete level by using
any of the locally conservative discretizations for the convective and pressure terms.
The case in which the divergence terms $\mC_{\rho E}^D$ and $\mP_{\rho E}^D$
(see Eq.\eqref{eq:P_rhoE} below)
are used is the simplest, and gives a locally conservative
expression for the total-energy equation that guarantees
that discrete total energy is preserved both globally and locally.
However, for the discretization of the convective term any linear combination of the forms in Eq.~\eqref{eq:Mom_Forms} is fine
for local conservation of total energy, independently on the choice of the
discretization of the mass and momentum fluxes.
It is customary, in absence of additional indications,
to use a
\revtwo{generalized kinetic-energy preserving discretization of the type in Eq.~\eqref{eq:Cdecomp} also for the convective terms in the total-energy equation (i.e. by assuming $\phi = E$).}
In this case the convective flux is given (in the second-order case)
by $\mC_{\rho E}=\llbracket\overline{E}\,m_{i+1/2}\rrbracket$, where $m_{i+1/2}$
is given by Eq.~\eqref{eq:MassFlux} and is the same mass flux adopted in the continuity and momentum equations.
This choice does not affect the local conservation of $\rho E$, but
gives the additional property that the quantity $\rho E^2$
is preserved by convection.
Experience shows that this additional structural property
has beneficial effects on the robustness of the simulation and,
among the various possible KEP forms, the KGP form ($\xi=1/2$) has shown
the best performances~\cite{Pirozzoli2010,Coppola2019}.
However, a KEP formulation for the  `energy' equation is not necessarily the best option,
as we will see in the subsequent sections.

The employment of central schemes gives a locally conservative structure also for
both the divergence and the advective forms of the pressure term:
\begin{equation}
    \mathcal{P}_{\rho E}^D=\dfrac{\delta pu_{\alpha}}{\delta x_{\alpha}}\;, \qquad
    \mathcal{P}_{\rho E}^A = p\dfrac{\delta u_{\alpha}}{\delta x_{\alpha}}+u_{\alpha}\dfrac{\delta p}{\delta x_{\alpha}}.\label{eq:P_rhoE}
\end{equation}
Although the use of the divergence form $\mP_{\rho E}^D$ seems the most natural choice, in principle one can use any (convex) linear combination of the divergence and advective forms,
without affecting local and global conservation of $\rho E$.
This means that the pressure term can be split as
\beq
\mP_{\rho E} = \chi\dfrac{\delta \rho \ua}{\delta \xa} +
\left(1-\chi\right)\left( p\dfrac{\delta \ua}{\delta\xa}+\ua\dfrac{\delta p}{\delta\xa}\right)
\eeq
with the corresponding finite volume formulation having flux
$\chi\,\overline{pu}+\left(1-\chi\right)\overline{\overline{\left(p,u\right)}}$.
Moreover, the pressure term can be written also in terms of the scaled pressure $\hat{p}=p/\rho$.
In this case it has the form of the convective term $\partial\rho\ua\hat{p}/\partial x_{\alpha}$ and any of the forms in
Eq.~\eqref{eq:Mom_Forms} (or any linear combination of them) can be used.
When for the scaled pressure the same splitting used for $\rho E$ is adopted, the pressure can be
included in the convective term for total energy,
obtaining a single convective term for the enthalpy $H = E+\hat{p}$.
This is the choice made by Jameson~\cite{Jameson2008b} and Pirozzoli~\cite{Pirozzoli2010}.
However, there is no a priori reason to consider a KEP splitting for
$\partial\rho\ua\hat{p}/\partial x_{\alpha}$, and different options could be more advantageous.

\subsubsection{Internal-energy equation}
The internal-energy equation has been used in many contributions as a `primary' energy variable in place of the equation for total energy, especially for LES studies, since the discretization of the internal-energy equation requires only modeling the SGS heat-flux term~\cite{SpyropoulosAIAA1996}.
Among the various studies employing different formulations of the internal energy equation, we mention here Moin \emph{et al.}~\cite{MoinPoF1991}, Blaisdell \emph{et al.}~\cite{Blaisdell1996} and Spyropoulos and Blaisdell~\cite{SpyropoulosAIAA1996}.
A recent paper by Veldman~\cite{Veldman2021} analyzes the compatibility relations
that the discrete terms have to satisfy to design a supraconservative
formulation when using the internal energy equation.

The discrete induced equation for internal energy is obtained by
using $G(\rho,u_{\alpha},E) = \rho e = \rho E - \rho u_{\alpha}u_{\alpha}/2$,
which gives the partial derivatives
$G_{\rho}=E-u_{\alpha}u_{\alpha}/2, G_{u_{\alpha}}=-\rho u_{\alpha}$ and $G_E = \rho$.
Eq.~\eqref{eq:Gevol} gives:
\beq\label{eq:rhoe_from_rhoE}
\frac{\partial\rho e}{\partial t} = -\underbrace{\left[\mC_{\rho E} - \underbrace{\left(u_{\alpha}\mC_{\rho u_{\alpha}}-\dfrac{u_{\alpha}u_{\alpha}}{2}\mM\right)}_{\mC_{\rho \kappa}} \right]}_{\mC_{\rho e}} -  \underbrace{\left(\mP_{\rho E}-\underbrace{u_{\alpha}\mP_{\rho u_{\alpha}}}_{\mP_{\rho \kappa}}\right)}_{\mP_{\rho e}},
\eeq
which could have been derived also by subtracting Eq.~\eqref{eq:KinEnergy_a} to the discrete equation for total energy.
Equation \eqref{eq:rhoe_from_rhoE} shows that the discrete evolution of the internal energy is driven by a convective term $\mC_{\rho e}$ obtained as the difference
between the convective term in the total-energy equation and that
in the induced equation for kinetic energy.
When the equation for total energy is directly discretized (together with mass and momentum equations), the terms
$\mC_{\rho E}, \mC_{\rho \ua}, \mM, \mP_{\rho E}$ and $\mP_{\rho \ua}$
in Eq.~\eqref{eq:rhoe_from_rhoE} are fixed by the discretization details, whereas
$\mC_{\rho e},\mC_{\rho\kappa}$ and $\mP_{\rho e}$ are the induced
discretizations in the implicit evolution equation for $\rho e$.
Since we assumed that the equation for mass and momentum are discretized
with a KEP scheme and the convective term in the discrete equation for $\rho E$ is
in locally-conservative form, both the terms $\mC_{\rho E}$ and $\mC_{\rho \kappa}$
are locally conservative, which implies that the convective term in the equation for $\rho e$ is automatically in locally conservative form.
To be concrete, if $\mathcal{F}_{\rho E}$ is the total-energy flux (which is fixed by the discretization; explicitly in the case of FV, or implicitly in the case of FD),
the convective term in the total-energy equation can be written as
$\mC_{\rho E} = \left\llbracket \mathcal{F}_{\rho E}\right\rrbracket$.
In Sec.~\ref{sec:LocalConservation} we saw that $\mC_{\rho \kappa}$ is
in locally-conservative form when a KEP discretization for momentum is adopted. In particular, for
a second-order discretization, it is expressed as $\mC_{\rho\kappa} = \left\llbracket{u_iu_{i+1}}m_{i+1/2}/{2}\right\rrbracket.$
The convective term for the induced internal-energy equation can be hence written as:
\beq\label{eq:Ce_Flux_TotE}
\mC_{\rho e} = \left\llbracket\ \mathcal{F}_{\rho E}-\frac{u_iu_{i+1}}{2}m_{i+1/2}\right\rrbracket,
\eeq
showing that $\rho e$ is preserved by convection both locally and globally.
In the particular case in which the equation for $\rho E$ is also discretized with a KEP scheme,
the most general form of $\mF_{\rho E}$ is, for second order discretizations,
$\mF_{\rho E} = \overline{E}m_{i+1/2}$
(with $m_{i+1/2}$ the same mass flux used for mass and momentum equations),
which implies
\beq\label{eq:Ce_Flux_TotE_KEP}
\mC_{\rho e} = \left\llbracket\left(\overline{E} -
\dfrac{u_iu_{i+1}}{2}\right)m_{i+1/2}\right\rrbracket.
\eeq
Equation~\eqref{eq:Ce_Flux_TotE_KEP} shows that the convective term in the induced equation for the internal energy is in locally conservative form, but not in KEP form, since in general the difference
$\overline{E}-u_iu_{i+1}/2$ cannot be cast as the
arithmetic average $\overline{e}$, which is the necessary and sufficient condition for second-order KEP fluxes~\cite{Veldman2019}.
This means that $\rho e$ is locally conserved, but $\rho e^2$ is not.

Equation~\eqref{eq:rhoe_from_rhoE} shows the relation between the discrete convective and pressure terms in
total and internal energies equations.
To better comment on this relation we enrich our notation by denoting with
bold characters the convective terms that are directly discretized
(e.g. $\boldsymbol{\mM}, \boldsymbol{\mC}_{\rho \ua}, \boldsymbol{\mC}_{\rho E}$ or $\boldsymbol{\mP}_{\rho E}$), whereas the convective terms in the induced discrete equations are denoted with the usual symbols
(e.g. $\mC_{\rho \kappa}$ or $\mC_{\rho e}$).
With this notation we can express the induced convective and pressure terms for the internal energy as functions of the directly discretized terms in mass, momentum and total energy with the relations:
\begin{equation}\label{eq:rhoe_induced_rhoE}
\begin{array}{ll}
     \mC_{\rho e} &= \boldsymbol{\mC}_{\rho E}- \left(u_{\alpha}\boldsymbol{\mC}_{\rho u_{\alpha}}
     -\dfrac{\ua\ua}{2}\boldsymbol{\mM}\right)= \boldsymbol{\mC}_{\rho E} -\mC_{\rho \kappa}  \\
     \mP_{\rho e} &= \boldsymbol{\mP}_{\rho E} - \ua\boldsymbol{\mP}_{\rho\ua}.
\end{array}
\end{equation}
However, it is evident that these relations
are valid independently of which equation is directly discretized, and can be inverted in order to express
$\mC_{\rho E}$ and $\mP_{\rho E}$ as functions of the convective and pressure terms for mass, momentum and internal energy.
This inversion is useful if one wants to express the induced evolution of total energy when the internal-energy equation is directly discretized.
In fact, a direct discretization of mass, momentum and internal energy equations settles these three quantities as
`primary' variables, whereas total energy is a secondary variable, whose evolution
is determined by an  induced discrete equation.
The evolution equation for the total energy in this case is governed by convective and pressure terms which are given by:
\begin{align}
     \mC_{\rho E} &=  \boldsymbol{\mC}_{\rho e}+ \underbrace{\left(u_{\alpha}\boldsymbol{\mC}_{\rho u_{\alpha}}
     -\dfrac{\ua\ua}{2}\boldsymbol{\mM}\right)}_{\mC_{\rho\kappa}}\label{eq:CrhoE_induced_rhoe}  \\
     \mP_{\rho E} &= \boldsymbol{\mP}_{\rho e} + \ua\boldsymbol{\mP}_{\rho\ua}.\label{eq:PrhoE_induced_rhoe}
\end{align}
Again, Eq.~\eqref{eq:CrhoE_induced_rhoe}-\eqref{eq:PrhoE_induced_rhoe} shows that when the internal energy is directly discretized with a locally conservative formulation, the convective term in the induced total-energy equation is also in locally conservative form, but its discretization is not in KEP form.
In this case, the equations analogous to Eq.~\eqref{eq:Ce_Flux_TotE}
and \eqref{eq:Ce_Flux_TotE_KEP} are
\beq\label{eq:CE_Flux_rhoe}
\mC_{\rho E} = \left\llbracket\mF_{\rho e} +
\dfrac{u_iu_{i+1}}{2}m_{i+1/2}\right\rrbracket
\eeq
\beq\label{eq:CE_Flux_rhoe_KEP}
\mC_{\rho E} = \left\llbracket\left(\overline{e} +
\dfrac{u_iu_{i+1}}{2}\right)m_{i+1/2}\right\rrbracket.
\eeq

Note that the requirement that $\mC_{\rho e}$ is in the KEP form with local flux $\mF_{\rho e}=\overline{e}\,m_{i+1/2}$ (cfr. Eq.~\eqref{eq:CE_Flux_rhoe_KEP})
can be expressed as the fact that the discrete convective operator for $\rho e$ is the same as the convective operator for momentum, which is $\mC_{\rho \ua} = \llbracket\overline{\ua}\,m_{i+1/2}\rrbracket$. This is an expression of the Requirement 3.3 stated by Veldman~\cite{Veldman2021}, which is a sufficient, but not necessary, condition for global conservation of total energy.

The pressure term in Eq.~\eqref{eq:PrhoE_induced_rhoe} is expressed as the sum of the pressure terms in the internal-energy equation $p\delta u/\delta x$ and the term $u\mP_{\rho u}=u\delta p/\delta x$.
This term is in locally conservative form if, and only if, it constitutes an advective form for the derivative of the product $pu$, which means that the two derivative matrices
acting on $u$ and $p$ satisfy the relation
\beq\label{eq:CondGradP}
{\sf D}_p = - {\sf D}_u^T
\eeq
which is always satisfied in our case, since we use the same (skew-symmetric) derivative operator in
$\mP_{\rho e}$ and $\mP_{\rho u}$.
The condition in Eq.~\eqref{eq:CondGradP} coincides with the Requirement 3.2 in the paper by Veldman~\cite{Veldman2021} (cfr. Eq.~(3.6) in \cite{Veldman2021}).

As a final remark, we explicitly note that, assuming exact time integration, it is actually immaterial which equation is directly discretized, since only the specification of the discrete terms $\mC$ and $\mP$ matters.
As an example, one could directly discretize the internal-energy equation with convective fluxes given by Eq.~\eqref{eq:Ce_Flux_TotE}
and the pressure term given by $\mP_{\rho e} = \mP_{\rho E}-\ua\mP_{\rho \ua}$.
In this case, given a KEP discretization for mass and momentum equations with the same mass flux, the numerical results
will be identical (for exact time integration) to
a discretization of the $\rho E$ equation with convective flux $\mF_{\rho E}$
and pressure term $\mP_{\rho E}$.

In the literature it is often encountered the case in which a certain formulation is expressed as a direct discretization of the total-energy equation with some complex expression for the convective terms or fluxes.
In many cases this happens because most of the existing codes are written by using a direct discretization of the total-energy equation.
In this case, new ideas are more straightforwardly implemented
by specifying the fluxes for the total energy, even if they could be more neatly expressed as fluxes for other thermodynamic variables.
This is the case, for example, of the entropy-preserving scheme by Honein and Moin~\cite{HoneinMoin2004}, who report a complex discrete equation for total energy emulating a direct discretization of the entropy equation with a Feiereisen form (Eq.~(19) in \cite{HoneinMoin2004}. See also next Sec.~\ref{sec:Entropy}).
In some other cases the convective fluxes in the total-energy equation are specified according to some physical or mathematical requirement, and the equivalence of the resulting formulation with others
involving different variables is not evident at first sight.
An example of this situation is given by the family of KEEP schemes discussed by Kuya \emph{et al.}~\cite{Kuya2018}.
This class of schemes are constructed by specifying convective and pressure terms according to the so-called \emph{Analytical relations}, which dictate
the form of the fluxes for the convective and pressure terms in the total-energy equation which can be expressed, in our notation, as (compare to Eqs. (15), (21), (33), (41) and (50) of \cite{Kuya2018}):
\beq
\begin{array}{ll}
     m_{i+1/2} &= \overline{\rho}\,\overline{u}  \\
     \mF_{\rho u} &= \overline{u}\, m_{i+1/2} \\
     \mF_{\rho E} &= \mF_{\rho e} + \mF_{\rho \kappa} \\
     \mP_{\rho E} &= \llbracket\overline{\overline{\left(p,u\right)}}\rrbracket =
     \mP_{\rho e} + \mP_{\rho\kappa}\\
\end{array}
\eeq
with
\beq
     \mF_{\rho e}  = \overline{e}\,m_{i+1/2},  \quad
     \mF_{\rho\kappa} = \dfrac{u_iu_{i+1}}{2}m_{i+1/2}, \quad
     \mP_{\rho e}  = p\dfrac{\delta u}{\delta x},  \quad
     \mP_{\rho\kappa} = u\dfrac{\delta p}{\delta x}.
\eeq
According to the discussion presented in the first part of this section, this formulation is equivalent (for exact time integration) to a direct discretization of the internal energy equation with a KEP scheme employing the KGP ($\xi = 1/2$) form.
This class of schemes has shown to be quite robust and with good properties of entropy conservation for the inviscid Taylor-Green flow, although they are not strictly conservative of entropy, as it is shown in Sec.~\ref{sec:Entropy}.
These  results were confirmed also in~\cite{Coppola2019}, where the equivalent scheme, formulated in terms of internal energy, is analyzed as the KGP$\left({\rho e}\right)$ formulation.

\subsubsection{Pressure equation}
\label{sec:Pressure}
The relation $p=\left(\gamma-1\right)\rho e$, which is valid for a perfect gas,
shows that the discrete induced equation for the pressure $p$
follows immediately from Eq.~\eqref{eq:rhoe_from_rhoE}.
The discrete convective and pressure terms in the pressure equation are simply proportional
to  $\mC_{\rho e}$ and $\mP_{\rho e}$ and the induced discrete equation for $p$
inherits all the properties of the discrete equation for $\rho e$.

The pressure equation has been occasionally used in previous works as the primary energy variable, the most notable example being the classical paper by Feiereisen \emph{et al.}~\cite{Feiereisen1981}.
When the pressure equation is directly discretized, the discrete properties of the induced equations for the other `energy' variables are similar to that
obtained in the case in which the internal energy is directly discretized.
The main difference between the two formulations (pressure and internal energy) is that, since the convective term in the pressure equation is constituted by the derivative of the product between  $p$ and $u$, the possible splittings which can be adopted are similar to that of the continuity equation, i.e.~the divergence $\mC^D_{p} = \delta pu/\delta x$ or
advective $\mC^A_{p} =p \delta u/\delta x+u\delta p/\delta x$ forms.
These forms correspond to the two particular splittings in the internal energy equation $\mC^D_{\rho e}$ and $\mC_{\rho e}^{u}$, in which the internal energy flux $(\rho u e)$ is split as the product of $\rho e$ and $u$ and not,
as it is usual, as the product of the mass flux $\rho u$ and $e$.
A splitting obtained by only using the forms $\mC^D_{\rho e}$ and $\mC^u_{\rho e}$
cannot be KEP, which means that it cannot preserve $\rho e^2$.
However, since it corresponds to a locally conservative discretization for the convective term in the pressure equation, it possesses some interesting properties, the most interesting one being the so-called Pressure Equilibrium Preservation (PEP) property, which has been the subject of recent studies by several authors~\cite{Shima2021}, \cite{Singh2021}, \cite{Ranocha2021}.

A PEP formulation is a discrete formulation of the balance equations that is able to preserve the equilibrium of velocity and pressure when they are constant at the initial time.
In fact, in this particular case the evolution equation for the pressure (Eq.~\eqref{eq:Pressure}) and the equation for the velocity $u$:
\beq\label{eq:Velocity}
\dfrac{\partial u}{\partial t} = -u\dfrac{\partial u}{\partial x} - \dfrac{1}{\rho}\dfrac{\partial p}{\partial x}
\eeq
predict that the initial constant state remains constant during the evolution,
as the right hand sides of Eq.~\eqref{eq:Pressure} and \eqref{eq:Velocity}
are both zero for constant $u$ and $p$.
A formulation that is able to reproduce this property at discrete level is termed PEP.

To investigate this class of discretizations we write the induced discrete version of Eq.~\eqref{eq:Velocity} by using Eq.~\eqref{eq:Gevol} with $G(u) = u, G_{\rho} = G_{\phi} = 0$ and $G_u = 1$, which gives
\beq\label{eq:DiscrVelEvol}
\dfrac{\partial u}{\partial t} = \dfrac{u}{\rho} \mM-\dfrac{\mC_{\rho u}}{\rho}
-\dfrac{\mP_{\rho u}}{\rho}.
\eeq
Assuming $u$ and $\rho$ constant, the discrete reproduction of the PEP property requires that the right hand side of Eq.~\eqref{eq:DiscrVelEvol} is zero.
It is readily seen that \authors{a direct} discretization of the pressure term $\mP_{\rho u} = \authors{\delta p/\delta x}$ is zero for spatially constant $p$.
This implies that a necessary condition for PEP is that for constant $u$ one has
\beq\label{eq:PEP_Cond_1}
u\mM = \mC_{\rho u}.
\eeq
Eq.~\eqref{eq:PEP_Cond_1} is automatically satisfied by any KEP formulation, since in this case one has $\mC_{\rho u} = \llbracket\overline{u}\,m_{i+1/2}\rrbracket$,
which implies Eq.~\eqref{eq:PEP_Cond_1} for constant $u$.
Note that a condition similar to Eq.~\eqref{eq:PEP_Cond_1} has been derived by Ranocha~\cite{Ranocha2021} in the framework of FV methods.

In addition to Eq.~\eqref{eq:PEP_Cond_1} a PEP formulation should also discretely satisfy $\partial p/\partial t = 0$ for constant $u$ and $p$, which implies
\beq\label{eq:PEP_Cond_2}
\mC_p + \mP_{p}= 0
\eeq
for constant $u$ and $p$.
This condition has to be checked case by case depending on the energy variable discretized and on the formulation adopted.
Any direct discretization of the pressure equation satisfies Eq.~\eqref{eq:PEP_Cond_2}, since in this case one has
\begin{align}
\mC_p &= \chi\dfrac{\delta pu}{\delta x} + \left(1-\chi\right)\left(p\dfrac{\delta u}{\delta x} +u\dfrac{\delta p}{\delta x} \right) \label{eq:Conv_p_discr}\\
\mP_p &= \left(\gamma-1\right)p\dfrac{\delta u}{\delta x}
\end{align}
which are readily seen to be individually zero for constant $u$ and $p$.
Starting from the discretization of the internal energy equation, one has
$\mC_{p}=\left(\gamma -1\right)\mC_{\rho e}$ and the condition
\eqref{eq:PEP_Cond_2} is satisfied when the splitting of the term $\partial\rho u e/\partial x$
is made by averaging only $\mC_{\rho e}^D$ and $\mC_{\rho e}^u$:
\beq \label{eq:PEP_Cond_rhoe}
\mC_{\rho e} = \chi\left(\dfrac{\delta \rho ue}{\delta x}\right)+
\left(1-\chi\right)
\left(u\dfrac{\delta \rho e}{\delta x} +
\rho e\dfrac{\delta u}{\delta x}\right)
\eeq
which corresponds to Eq.~\eqref{eq:Conv_p_discr}.
This last splitting has been adopted by Shima \emph{et al.}~\cite{Shima2021}, who used the value $\chi = 1/2$.

A direct discretization of the total energy equation gives induced
$\mC_p$ and $\mP_p$ terms:
\begin{align}
    \mC_p &= \left(\gamma-1\right)\left(\mC_{\rho E}-\mC_{\rho\kappa}\right)\label{eq:Cp_from_rhoE}\\
    \mP_p &= \left(\gamma-1\right)\left(\mP_{\rho E}-u\mP_{\rho u}\right).\label{eq:Pp_from_rhoE}
\end{align}
A straightforward discretization of the term $\partial p/\partial x$ in the momentum equation typically leads to $\mP_{\rho u}=0$ for constant $p$.
In this case plugging Eq.~\eqref{eq:Cp_from_rhoE} and \eqref{eq:Pp_from_rhoE} into  Eq.~\eqref{eq:PEP_Cond_2} requires
\beq\label{eq:PEP_Cond_2_rhoE}
\mC_{\rho E}-\mC_{\rho \kappa} +\mP_{\rho E}= 0
\eeq
for  constant $u$ and $p$.
This condition can be satisfied in several ways.
As an example, one can split the term $\mC_{\rho E}$ in its contributions
due to internal and kinetic energies: $\mC_{\rho E}=\mC_{\rho E}^{\rho e}+\mC_{\rho E}^{\rho \kappa}$. For constant $u$ and $p$ any discretization of the term $\mC_{\rho E}^{\rho \kappa}$ from the family of forms in Eq.~\eqref{eq:Mom_Forms} reduces to $(u^3/2)\delta\rho/\delta x$,
as the induced discrete term $\mC_{\rho\kappa}$ in Eq.~\eqref{eq:rhoe_from_rhoE} does.
Equation~\eqref{eq:PEP_Cond_2_rhoE} hence dictates $\mC_{\rho E}^{\rho e}+\mP_{\rho E} = 0$.
Since the convective term for $\rho e$ can be reduced to a convective term for $p$ by using the relation $\rho e = p/(\gamma - 1)$,
one is left with the condition that the
discretization of the term $(\gamma/(\gamma-1))\partial p u/\partial x$
should reduce to zero for constant $u$ and $p$, which is a condition satisfied by any linear combination of the forms in Eq.~\eqref{eq:P_rhoE}.
In a recent paper \cite{Singh2021} Singh and Chandrashekar proposed a new KEP
scheme, which is also PEP, based on this approach, for which they use a KGP form for $\mC_{\rho E}^{\rho \kappa}$ and an arithmetic average of the forms in Eq.~\eqref{eq:P_rhoE} for the pressure term coming from $\mC_{\rho E}^{\rho e}$ and $\mP_{\rho E}$.

\subsubsection{Total enthalpy equation}\label{sec:Enthalpy}
Total enthalpy $\rho H$ is the sum of total energy $\rho E$ and pressure $p$.
Its discrete evolution equation can be hence obtained by simply summing the
discrete equations for $\rho E$ and $p$.
The convective and pressure terms for the total enthalpy, expressed as functions
of the discrete terms in the equation for total energy, are given by
\begin{align}
\mC_{\rho H} &= \boldsymbol{\mC}_{\rho E} + (\gamma -1)\left(\boldsymbol{\mC}_{\rho E} -\mC_{\rho \kappa}\right)\label{eq:Conv_Enthalpy_rhoE}\\
\mP_{\rho H} &= \boldsymbol{\mP}_{\rho E} + (\gamma -1)\left(\boldsymbol{\mP}_{\rho E}-\ua\boldsymbol{\mP}_{\ua}\right)\label{eq:Pres_Enthalpy_rhoE}
\end{align}
from which one can easily conclude that for KEP discretization of momentum equations
and locally conservative discretizations of the total-energy equation, the convective term in the total-enthalpy equation is in locally conservative form, with flux given by
\beq\label{eq:Flux_Enthalpy_rhoE}
\mF_{\rho H}=\gamma\mF_{\rho E}-\left(\gamma-1\right)\dfrac{u_iu_{i+1}}{2}m_{i+1/2}.
\eeq
When starting from a direct discretization of the internal-energy equation, by substituting Eq.~\eqref{eq:CrhoE_induced_rhoe} into Eq.~\eqref{eq:Conv_Enthalpy_rhoE}--\eqref{eq:Flux_Enthalpy_rhoE}
one has
\begin{align}
\mC_{\rho H} &= \gamma\boldsymbol{\mC}_{\rho e} + \mC_{\rho \kappa}\label{eq:Conv_Enthalpy_rhoe}\\
\mP_{\rho H} &= \gamma\boldsymbol{\mP}_{\rho e} +  \ua\boldsymbol{\mP}_{\ua}\label{eq:Pres_Enthalpy_rhoe}\\
\mF_{\rho H} &= \gamma\mF_{\rho e}+\dfrac{u_iu_{i+1}}{2}m_{i+1/2}.\label{eq:Flux_Enthalpy_rhoe}
\end{align}
As usual, Eq.~\eqref{eq:Conv_Enthalpy_rhoE}-\eqref{eq:Flux_Enthalpy_rhoE} or
\eqref{eq:Conv_Enthalpy_rhoe}-\eqref{eq:Flux_Enthalpy_rhoe}
can be inverted to obtain the convective term (or its local flux) of the induced total-energy or internal-energy equations when the total-enthalpy equation is directly discretized as a primary variable.
The relations
\begin{align}
\mC_{\rho E} &= \dfrac{1}{\gamma}\boldsymbol{\mC}_{\rho H} + \left(\dfrac{\gamma-1}{\gamma}\right)\mC_{\rho \kappa}\label{eq:Conv_TotEnergy_rhoH}\\
\mC_{\rho e} &= \dfrac{1}{\gamma}\left(\boldsymbol{\mC}_{\rho H} \authors{-} \mC_{\rho \kappa}\right)\label{eq:Conv_IntEnergy_rhoH}
\end{align}
show that when discretizing the total-enthalpy equation with a locally conservative formulation, also internal and total energies are preserved by convection
(provided a KEP formulation is adopted for momentum).
The pressure terms are analogously given by
\begin{align}
\mP_{\rho E} &= \dfrac{1}{\gamma}\boldsymbol{\mP}_{\rho H} + \left(\dfrac{\gamma-1}{\gamma}\right)u\boldsymbol{\mP}_{\rho u}\label{eq:Pres_TotEnergy_rhoH}\\
\mP_{\rho e} &= \dfrac{1}{\gamma}\left(\boldsymbol{\mP}_{\rho H} \authors{-} u\boldsymbol{\mP}_{\rho u}\right).\label{eq:Pres_IntEnergy_rhoH}
\end{align}
Note that, since
\beq\label{eq:PresTotEnthalpy}
\mP_{\rho H} = \dfrac{\delta pu}{\delta x} - \left(\gamma - 1\right)p\dfrac{\delta u}{\delta x},
\eeq
a straightforward  discretization of the non-conservative term in Eq.~\eqref{eq:PresTotEnthalpy} with a discrete derivative related to that in momentum equation by Eq.~\eqref{eq:CondGradP} guarantees that the term $\mP_{\rho E}$ from Eq.~\eqref{eq:Pres_TotEnergy_rhoH} is in locally conservative form.

As for the case of the total energy, the term $\partial pu/\partial x$
contained in $\mP_{\rho H}$  can be expressed by using the scaled pressure $\hat{p}$ as $\partial u\rho\hat{p}/\partial x$, for which a triple product splitting can be used.
As for the Jameson-Pirozzoli formulation, it can be discretized by using the same splitting of $\mC_{\rho H}$, which means that the conservative part of the pressure term $\mP_{\rho H}$ can be included into the convective term for $\rho H$ to have a single convective term for the quantity $\rho H + p$.
Alternatively, the pressure component of the convective term $\rho H = \rho E+p$ can be separately discretized together with the conservative part of $\mP_{\rho H}$.

\subsubsection{Sound speed equation} \label{sec:SoundSpeed}

The discrete evolution equation for the sound speed $\rho c$ can be
derived from the equation for $\rho e$ by using Eq.~\eqref{eq:Gevol}
applied to the relation $\rho c = G(\rho e)$.
In the case of a perfect gas one has  $c^2=e\gamma(\gamma-1) $
and Eq.~\eqref{eq:Gevol} furnishes
\beq\label{eq:rhoc_from_rhoe}
\dfrac{\partial \rho c}{\partial t} = -\dfrac{c}{2}\mM-
\dfrac{\gamma(\gamma-1)}{2c}\left(\mC_{\rho e}+\mP_{\rho e}\right).
\eeq
Equation~\eqref{eq:rhoc_from_rhoe} expresses the fact that a direct discretization
of the equation for $\rho e$ induces convective and pressure terms in the discrete equation for $\rho c$ given by
\begin{align}
     \mC_{\rho c} &= \dfrac{\gamma(\gamma-1)}{2c}\boldsymbol{\mC}_{\rho e} +
                     \dfrac{c}{2}\boldsymbol{\mM}\label{eq:Cc_induced_rhoe}\\
     \mP_{\rho c} &= \dfrac{\gamma(\gamma-1)}{2c}\boldsymbol{\mP}_{\rho e}.\label{eq:Pc_induced_rhoe}
\end{align}
Of course, the expression of $\mC_{\rho c}$ and $\mP_{\rho c}$
as functions of the discrete terms in the total energy equation,
which are relevant when the equation for $\rho E$ is directly discretized, is obtained by substituting Eq.~\eqref{eq:CrhoE_induced_rhoe} and \eqref{eq:PrhoE_induced_rhoe}  into Eq.~\eqref{eq:Cc_induced_rhoe}--\eqref{eq:Pc_induced_rhoe}.

Equations~\eqref{eq:Cc_induced_rhoe}--\eqref{eq:Pc_induced_rhoe} (and their counterpart for $\rho E$) show that when
the internal (or total) energy is directly discretized, the induced sound-speed convection term is,
in general, not in conservative form, indicating that the discrete evolution of $\rho c$ can be affected by spurious production from the discrete convective terms.
However, inversion of Eq.~\eqref{eq:Cc_induced_rhoe} and \eqref{eq:Pc_induced_rhoe} gives:
\begin{align}\label{eq:rhoe_induced_rhoc}
     \mC_{\rho e} &= \dfrac{2}{\gamma(\gamma-1)}\left(c\,\boldsymbol{\mC}_{\rho c} -\dfrac{c^2}{2}\boldsymbol{\mM}\right) \\
     \mP_{\rho e} &= \dfrac{2c}{\gamma(\gamma-1)}\boldsymbol{\mP}_{\rho c}
\end{align}
from which one can note that the convective term for the internal-energy equation,
corresponding with a discretization of the sound-speed equation,
has the basic structure of that of a generalized kinetic energy (cfr. Eq.~\eqref{eq:GenKinEnergy_a}).
This shows that a 
KEP discretization of the sound-speed equation
induces a locally conservative discretization for $\rho e$ with local flux
$\sqrt{e_i e_{i+1}}\,m_{i+1/2}$, which in turn implies a conservative
induced discretization also of $\rho E$, when a KEP scheme is used also for mass and momentum.
This observation was already made by Kok~\cite{Kok2009}, who used the sound speed as
the primary energy variable to derive a conservative approximation for both $\rho e$ and $\rho E$.
In his paper, Kok uses a Feiereisen splitting for the sound-speed convection term.
In the Numerical Results Section we show how the use of a KGP splitting can greatly improve the robustness
and the conservation properties of this formulation.

The discussion made in Sec.~\ref{sec:Discrete_KEP_evol} suggests that
a direct discretization of the $\rho e$ equation according to the splitting in Eq.~\eqref{eq:ConvKinGlobPres}
(with $\phi = \sqrt{2e}$) induces a locally conservative (and KEP) discretization of the sound-speed
convective term,
which can be equivalently expressed by saying that a splitting of the $\mC_{\rho e}$ term made according to Eq.~\eqref{eq:ConvKinGlobPres} furnishes a locally conservative form to $\mC_{\rho c}$ from Eq.~\eqref{eq:Cc_induced_rhoe}, with local flux $\mF_{\rho c} = \overline{c}\,m_{i+1/2}$.
When discretizing the internal energy equation, the splitting in Eq.~\eqref{eq:ConvKinGlobPres}
can be used in place of a classical KEP splitting (preserving $\rho e^2$) in order
to preserve $\rho c$ (proportional to $\rho \sqrt{e}$).
Total energy is of course also globally and locally conserved, with local flux
\beq
\mF_{\rho E} = \left(\sqrt{e_i e_{i+1}}+\dfrac{u_iu_{i+1}}{2}\right)m_{i+1/2}.
\eeq

\subsubsection{Entropy equation} \label{sec:Entropy}
The discrete equation for the entropy $\rho s$ as a function of $\rho e$ is
obtained by considering $G(\rho,e) = c_v\ln\left(p/\rho^{\gamma}\right)$,
for which one has $G_{\rho} = s+c_v(1-\gamma)$, $G_e = c_v\rho/e$. Eq.~\eqref{eq:Gevol}
gives:
\beq
\dfrac{\partial \rho s}{\partial t} = -\left(s-\gamma c_v\right)\mM - \dfrac{c_v}{e}\left(\mC_{\rho e}+\mP_{\rho e}\right).
\eeq
The relation expressing the convective term of the entropy equation as a function of the discrete terms in internal and total energy is:
\begin{align}
\mC_{\rho s} &= \left(s -\gamma c_v\right)\boldsymbol{\mM}\authors{+}\dfrac{c_v}{e}\left(\boldsymbol{\mC}_{\rho e} + \boldsymbol{\mP}_{\rho e}\right)\label{eq:Crhos_from_rhoE}, \\
\mC_{\rho s} &= \left(s -\gamma c_v\right)\boldsymbol{\mM}\authors{+}\dfrac{c_v}{e}\left(\boldsymbol{\mC}_{\rho E} - \mC_{\rho\kappa} + \boldsymbol{\mP}_{\rho E} - u\boldsymbol{\mP}_{\rho u}\right)\label{eq:Crhos_from_rhoe}.
\end{align}
To investigate if entropy is preserved by convection, one should consider if the term $\mC_{\rho s}$ in Eq.~\eqref{eq:Crhos_from_rhoe} or \eqref{eq:Crhos_from_rhoE} is in local (or global) conservation form.
This check is problematic in general and one has to adopt the approach to consider
the various forms of $\mC_{\rho s}$ for individual discretizations.
A careful analysis of the various discrete equations shows that in all the discretizations we have illustrated in the previous sections entropy is not strictly conserved,
because of the terms $s\mM$ and $c_v(\mC_{\rho e}+\mP_{\rho e})/e$, which cannot be cast, in general, as difference of fluxes. This circumstance is also confirmed by the numerical tests in Sec.~\ref{sec:NumRes}.
\revtwo{Note that this approach is basically equivalent to the classical entropy theory developed by Tadmor \cite{Tadmor1987,Tadmor2003} in the context of FV discretizations of systems of conservation laws admitting an entropy function. In fact, in the case of compressible Euler equations the mathematical entropy is given by $\eta = -\rho s$ and Tadmor's theory gives conditions for the existence of numerical fluxes for its induced discrete equation, starting from a FV discretization of mass, momentum and total energy. These conditions must be equivalent to the ones ensuring the locally conservative structure of the convective term $\mathcal{C}_{\rho s}$ we are seeking.}

\revtwo{The dual approach in which the entropy equation is directly discretized has been pursued by Honein and Moin~\cite{HoneinMoin2004}. In this case the entropy is a primary variable of the discretized system and its exact preservation is a direct consequence of the conservative structure of the discretization.
To investigate the conservation properties of this approach,} equations~\eqref{eq:Crhos_from_rhoE}-\eqref{eq:Crhos_from_rhoe} can be inverted, as usual,
to express the convective and pressure terms in the induced discrete equations for $\rho e$ and $\rho E$ when the entropy equation is directly discretized
\begin{align}
\mC_{\rho e} + \mP_{\rho e} &= \dfrac{e}{c_v}\boldsymbol{\mC}_{\rho s} - \dfrac{e}{c_v}\left(s -\gamma c_v\right)\boldsymbol{\mM}\label{eq:rhoe_from_rhos}\\
\mC_{\rho E} + \mP_{\rho E} &= \dfrac{e}{c_v}\boldsymbol{\mC}_{\rho s} - \dfrac{e}{c_v}\left(s -\gamma c_v\right)\boldsymbol{\mM} + \mC_{\rho\kappa} + u\boldsymbol{\mP}_{\rho u}\label{eq:rhoE_from_rhos}.
\end{align}
When substituting the Feiereisen form ($\xi = 1$) into mass, momentum and entropy equations  one has
\begin{equation}
\begin{array}{ll}
    \mM &=  \dfrac{\delta \rho u_{\alpha}}{\delta x_{\alpha}}, \\
    \mC_{\rho\kappa} &= \dfrac{1}{2}\left(u_{\beta}\dfrac{\delta\rho u_{\beta} u_{\alpha}}{\delta x_{\alpha}}+\rho u_{\beta} u_{\alpha}\dfrac{\delta u_{\beta}}{\delta x_{\alpha}}\right),\\
    \mC_{\rho s} &= \dfrac{1}{2}\left(\dfrac{\delta \rho u_{\alpha} s}{\delta x_{\alpha}} +
    \rho u_{\alpha}\dfrac{\delta s}{\delta x_{\alpha}} + s\dfrac{\delta \rho u_{\alpha}}{\delta x_{\alpha}}\right),\\
    \mP_{\rho \ua} &= \dfrac{\delta p}{\delta \xa}.
\end{array}
\end{equation}
and Eq.~\eqref{eq:rhoe_from_rhos} and \eqref{eq:rhoE_from_rhos} reduce to the nonviscous
version of the equations (18) and (19) reported by Honein and Moin~\cite{HoneinMoin2004}.
The application of a KGP splitting to this formulation has been investigated in \cite{Coppola2019} with reference to the
inviscid TGV flow and will be reconsidered here.

\begin{landscape}
\begin{table}
\renewcommand\arraystretch{1.8}
\centering
\revone{
\resizebox{!}{.28\linewidth}{%
\begin{tabular}{cc|cccccc}
&&\multicolumn{6}{c}{Primary variables}\\
&&$\rho E$  & $\rho e$ & $p$ & $\rho H$ & $\rho c$  & $\rho s$\\
\hline
\multirow{2}{*}{$\mR_{\rho E}$}  & $\mC_{\rho E}$  &  $\boldsymbol{\mC}_{\rho E} $ & $\boldsymbol{\mC}_{\rho e}+\mC_{\rho k} $ & $\dfrac{1}{\gamma-1}\boldsymbol{\mC}_{p}+\mC_{\rho k}  $ & $\dfrac{1}{\gamma}\boldsymbol{\mC}_{\rho H} + \dfrac{\gamma-1}{\gamma}\mC_{\rho k}$ & $\dfrac{\chat}{\gamma}\left(\,\boldsymbol{\mC}_{\rho c} -\dfrac{c}{2}\boldsymbol{\mM}\right) +\mC_ {\rho k}$ & \multirow{2}{*}{$\dfrac{e}{c_v}\boldsymbol{\mC}_{\rho s} - \dfrac{e}{c_v}\shat\boldsymbol{\mM} + \mR_{\rho k}$}\\
&$\mP_{\rho E}$  &  $\boldsymbol{\mP}_{\rho E} $ & $\boldsymbol{\mP}_{\rho e}+\mP_{\rho k}$& $\dfrac{1}{\gamma-1}\boldsymbol{\mP}_{p}+\mP_{\rho k} $ & $\dfrac{1}{\gamma}\boldsymbol{\mP}_{\rho H} + \dfrac{\gamma-1}{\gamma}\mP_{\rho k}$ & $\dfrac{\chat}{\gamma}\boldsymbol{\mP}_{\rho c}+\mP_ {\rho k}$\\
&\vspace*{-1.0em}\\
\multirow{2}{*}{$\mR_{\rho e}$}  &  $\mC_{\rho e}$  & $\boldsymbol{\mC}_{\rho E}- \mC_{\rho k} $ &$\boldsymbol{\mC}_{\rho e} $&$\dfrac{1}{\gamma-1}\boldsymbol{\mC}_{p} $&$\dfrac{1}{\gamma}\boldsymbol{\mC}_{\rho H} - \dfrac{1}{\gamma}\mC_{\rho k}$&   $\dfrac{\chat}{\gamma}\left(\boldsymbol{\mC}_{\rho c} -\dfrac{c}{2}\boldsymbol{\mM}\right)$ &            \multirow{2}{*}{
$\dfrac{e}{c_v}\boldsymbol{\mC}_{\rho s} - \dfrac{e}{c_v}\shat\boldsymbol{\mM}$} \\
&$\mP_{\rho e}$  &$\boldsymbol{\mP}_{\rho E} - \mP_{\rho k}$ &$\boldsymbol{\mP}_{\rho e} $&$\dfrac{1}{\gamma-1}\boldsymbol{\mP}_{p} $&$\dfrac{1}{\gamma}\boldsymbol{\mP}_{\rho H} - \dfrac{1}{\gamma}\mP_{\rho k}$ &  $\dfrac{\chat}{\gamma}\boldsymbol{\mP}_{\rho c}$
&\\
&\vspace*{-1.0em}\\
 \multirow{2}{*}{$\mR_{p}$}  & $\mC_{p}$  & $(\gamma-1)\left(\boldsymbol{\mC}_{\rho E}- \mC_{\rho k} \right)$ &$(\gamma-1)\boldsymbol{\mC}_{\rho e} $&$\boldsymbol{\mC}_{p}$&$\dfrac{\gamma -1}{\gamma}\boldsymbol{\mC}_{\rho H} - \dfrac{\gamma -1}{\gamma}\mC_{\rho k}$ & $\dfrac{2}{\gamma}\left(c\,\boldsymbol{\mC}_{\rho c} -\dfrac{c^2}{2}\boldsymbol{\mM}\right)$ &\multirow{2}{*}{
$\ehat\boldsymbol{\mC}_{\rho s} - \ehat\shat\boldsymbol{\mM}$}\\
&$\mP_{p}$  &$(\gamma-1)\left(\boldsymbol{\mP}_{\rho E} - \mP_{\rho k}\right)$ &$(\gamma-1)\boldsymbol{\mP}_{\rho e} $&$\boldsymbol{\mP}_{p}$ &$\dfrac{\gamma -1}{\gamma}\boldsymbol{\mP}_{\rho H} - \dfrac{\gamma -1}{\gamma}\mP_{\rho k}$&  $\dfrac{2c}{\gamma}\boldsymbol{\mP}_{\rho c}$
& \\
&\vspace*{-1.0em}\\
\multirow{2}{*}{$\mR_{\rho H}$}  &
 $\mC_{\rho H}$& $\gamma\boldsymbol{\mC}_{\rho E} - (\gamma-1)\mC_{\rho k}$ & $\gamma\boldsymbol{\mC}_{\rho e} + \mC_{\rho k}$ & $\dfrac{\gamma}{\gamma-1}\boldsymbol{\mC}_{p} + \mC_{\rho k}$ & $\boldsymbol{\mC}_{\rho H} $   &  $\chat\left(\boldsymbol{\mC}_{\rho c} -\dfrac{c}{2}\boldsymbol{\mM}\right) +\mC_ {\rho k}$  &      \multirow{2}{*}{
$\dfrac{\gamma e}{c_v}\boldsymbol{\mC}_{\rho s} - \dfrac{\gamma e}{c_v}\shat\boldsymbol{\mM}+ \mR_{\rho k}$}      \\
& $\mP_{\rho H}$& $\gamma\boldsymbol{\mP}_{\rho E} - (\gamma-1)\mP_{\rho k}$ & $\gamma\boldsymbol{\mP}_{\rho e} + \mP_{\rho k}$ & $\dfrac{\gamma}{\gamma-1}\boldsymbol{\mP}_{p} + \mP_{\rho k}$ & $\boldsymbol{\mP}_{\rho H} $   &  $\chat\boldsymbol{\mP}_{\rho c}+\mP_ {\rho k}$  \\
&\vspace*{-1.0em}\\
\multirow{2}{*}{$\mR_{\rho c}$}  &  $\mC_{\rho c}$ & $ \dfrac{\gamma}{\chat}\left(\boldsymbol{\mC}_{\rho E} - \mC_{\rho k}\right) +     \dfrac{c}{2}\boldsymbol{\mM} $ & $ \dfrac{\gamma}{\chat}\boldsymbol{\mC}_{\rho e} +     \dfrac{c}{2}\boldsymbol{\mM} $ &$ \dfrac{\gamma}{2c}\boldsymbol{\mC}_{p} +     \dfrac{c}{2}\boldsymbol{\mM} $ & $\dfrac{1}{\chat}\boldsymbol{\mC}_{\rho H} - \dfrac{1}{\chat}\mC_{\rho k} +     \dfrac{c}{2}\boldsymbol{\mM}$    & $\boldsymbol{\mC_{\rho c}} $ &       \multirow{2}{*}{$\dfrac{\gamma e}{\chat c_v}\boldsymbol{\mC}_{\rho s} - \dfrac{\gamma e}{\chat c_v}\shat\boldsymbol{\mM}+\dfrac{c}{2}\boldsymbol{\mM} $}\\
& $\mP_{\rho c}$ & $    \dfrac{\gamma}{\chat}\left(\boldsymbol{\mP}_{\rho E} - \mP_{\rho k}\right) $ &  $    \dfrac{\gamma}{\chat}\boldsymbol{\mP}_{\rho e} $    &$    \dfrac{\gamma}{2c}\boldsymbol{\mP}_{p} $ &  $\dfrac{1}{\chat}\boldsymbol{\mP}_{\rho H} - \dfrac{1}{\chat}\mP_{\rho k}$ & $\boldsymbol{\mP_{\rho c}} $ &      \\
&\vspace*{-1.0em}\\
$\mR_{\rho s}$  & $\mC_{\rho s}$ & $\!
    \shat\boldsymbol{\mM}+\dfrac{c_v}{e}\left(\boldsymbol{\mR}_{\rho E}- \mR_{\rho\kappa}\right)$
& $\shat\boldsymbol{\mM}+\dfrac{c_v}{e}\boldsymbol{\mR}_{\rho e}$
& $\shat\boldsymbol{\mM}+\dfrac{1}{\ehat}\boldsymbol{\mR}_{p}$
& $\shat\boldsymbol{\mM} +\dfrac{c_v}{\gamma e}\left(\boldsymbol{\mR}_{\rho H}- \mR_{\rho k}\right)$
& $\left(\shat - \dfrac{\chat \,c_v c}{2\gamma e}\right)\boldsymbol{\mM}+\dfrac{\chat\, c_v }{\gamma e}\boldsymbol{\mR}_{\rho c}$
& $\boldsymbol{\mC}_{\rho s}$\\
\hline
\end{tabular}
} 
} 
\caption{\revone{Summary of the different discretization options for the `energy' equation. On the columns, the convective $\mC$ and pressure $\mP$ terms induced by different choices of the primary variable. The bold characters indicate the terms that have been discretized directly.  When entropy is used as the primary variable it does not induce distinct convective and pressure terms for the other quantities, but only their sum $\mR = \mC+\mP$.The convective term $\boldsymbol{\mM}$ of the mass equation and the convective $\boldsymbol{\mC}_{\rho u_{\alpha}}$ and pressure $\boldsymbol{\mP}_{\rho u_{\alpha}}$ terms of the momentum equations are also directly discretized and, for the kinetic energy, induce $\mC_{\rho k} = u_{\alpha}\boldsymbol{\mC}_{\rho u_{\alpha}}
     -\dfrac{\ua\ua}{2}\boldsymbol{\mM}$ and $\mP_{\rho k} = u_{\alpha}\boldsymbol{\mP}_{\rho u_{\alpha}}$. The table uses the abbreviations $\shat = (s-\gamma c_v)$, $\ehat = \frac{e(\gamma-1)}{c_v}$, and $\chat = \frac{2c}{(\gamma-1)}$
     \label{tab:DiscretSummary}}}
\end{table}
\end{landscape}

 \begin{table}
\renewcommand\arraystretch{1.8}
\centering
\begin{tabular}{c|ccccccccc}
&&\multicolumn{2}{c}{Flux}&\multicolumn{6}{c}{Preserved variable}\\
&Ref.&$\mC$&$\mP$&   $\rho E$  & $\rho e$  & $\rho c$  & $\rho H$ & $\rho s$ & PEP\\
\hline
$(\rho E)$ &e.g. \cite{subbareddy2009} &$\llbracket \revone{\massflux} \,\overline{E}\rrbracket$&$\llbracket\overline{pu}\rrbracket$&  $\checkmark$ & $\checkmark$& $\times$ & $\checkmark$ & $\times$ & $\times$\\
$(\rho E)_{\text{JP}}$&\cite{Jameson2008b,Pirozzoli2010}&$\llbracket\revone{\massflux}\,\overline{E}\rrbracket$&$\llbracket\overline{\rho}\,\overline{u}\,\overline{\hat{p}}\rrbracket$& $\checkmark$ & $\checkmark$& $\times$ & $\checkmark$   & $\times$ & $\times$ \\
 $(\rho E)_{\text{PEP}}$ &\cite{Singh2021}&$\llbracket\overline{\rho e}\,\overline{u} + \revone{\massflux}\,\overline{\kappa}\rrbracket$&$\llbracket\overline{u}\,\overline{p}\rrbracket$& $\checkmark$ & $\checkmark$& $\times$ & $\checkmark$   & $\times$ & $\checkmark$    \\
\hline
 $(\rho e)$  &\cite{Kuya2018,Coppola2019}&$\llbracket\revone{\massflux}\,\overline{e}\rrbracket$&$p\frac{\delta u}{\delta x}$& $\checkmark$&$\checkmark$& $\times$& $\checkmark$   & $\times$ & $\times$    \\
  $(\rho e)_{\text{PEP}}$ &\cite{Shima2021}&$\llbracket\overline{\rho e}\,\overline{u}\rrbracket$&$p\frac{\delta u}{\delta x}$& $\checkmark$  & $\checkmark$& $\times$ & $\checkmark$   & $\times$ & $\checkmark$   \\
  $(\rho e)_{\text{div}}$ &new&$\llbracket\overline{\rho ue}\rrbracket$&$p\frac{\delta u}{\delta x}$& $\checkmark$ & $\checkmark$ & $\times$ & $\checkmark$& $\times$ & $\checkmark$    \\
\hline
 $(\rho c)_{\text{F}}$ &\cite{Kok2009}&$\llbracket\revone{\massflux_{\text{F}}}\,\overline{c}\rrbracket$&$\frac{\gamma -1}{2}\rho c\frac{\delta u}{\delta x}$& $\checkmark$ & $\checkmark$& $\checkmark$ & $\checkmark$   & $\times$ & $\times$    \\
  $(\rho c)_{\text{C}}$ &new&$\llbracket\revone{\massflux_{\text{C}}}\,\overline{c}\rrbracket$&$\frac{\gamma -1}{2}\rho c\frac{\delta u}{\delta x}$& $\checkmark$ & $\checkmark$& $\checkmark$ & $\checkmark$   & $\times$ & $\times$    \\
  $(\rho c)_{\text{KGP}}$ &new&$\llbracket\revone{\massflux_{\text{KGP}}} \,\overline{c}\rrbracket$&$\frac{\gamma -1}{2}\rho c\frac{\delta u}{\delta x}$& $\checkmark$ & $\checkmark$& $\checkmark$ & $\checkmark$   & $\times$ & $\times$    \\
\hline
  $(\rho H)$&new&$\llbracket\revone{\massflux} \,\overline{H}\rrbracket$&$\llbracket\overline{pu}\rrbracket-(\gamma-1)p\frac{\delta u}{\delta x}$& $\checkmark$ & $\checkmark$ & $\times$ & $\checkmark$   & $\times$ & $\times$     \\
$(\rho H)_{H+p}$ &new&$\llbracket\revone{\massflux} \,\overline{H}\rrbracket$&$\llbracket\overline{\rho}\,\overline{u}\overline{\hat{p}}\rrbracket-(\gamma-1)p\frac{\delta u}{\delta x}$& $\checkmark$ & $\checkmark$ & $\times$ & $\checkmark$   & $\times$ & $\times$     \\
  $(\rho H)_E$ &new&$\llbracket\revone{\massflux} \,\overline{E}+\overline{pu}\rrbracket$&$\llbracket\overline{pu}\rrbracket-(\gamma-1)p\frac{\delta u}{\delta x}$&$\checkmark$ & $\checkmark$ & $\times$ & $\checkmark$   & $\times$ & $\times$     \\
\hline
$(\rho s)_{\text{F}}$ &\cite{HoneinMoin2004}&$\llbracket\revone{\massflux_{\text{F}}} \,\overline{s}\rrbracket$&---& $\times$ & $\times$& $\times$ & $\times$   & $\checkmark$ & $\times$    \\
  $(\rho s)_{\text{C}}$ &\cite{Coppola2019}&$\llbracket\revone{\massflux_{\text{C}}} \,\overline{s}\rrbracket$&---& $\times$ & $\times$& $\times$ & $\times$   & $\checkmark$ & $\times$    \\
  $(\rho s)_{\text{KGP}}$ &\cite{Coppola2019}&$\llbracket\revone{\massflux_{\text{KGP}}} \,\overline{s}\rrbracket$&---& $\times$ & $\times$& $\times$ & $\times$   & $\checkmark$ & $\times$ \\
\hline
\end{tabular}
\caption{Fluxes and conservation properties of the various formulations considered.
Continuity and momentum equations \revtwo{have convective and pressure terms $\mM = \llbracket \massflux \rrbracket$, $\mC_{\rho u} = \llbracket \massflux \overline{u} \rrbracket$ and $\mP_{\rho u} = \delta p/\delta x$ ($\massflux$ is the mass flux); they are discretized with a KEP form, implying that mass, momentum and kinetic energy are always preserved}. \revone{With the F form, the mass flux $\massflux$ is discretized as $\massflux_{\text{F}} = \overline{\rho u}$; for the C form it is $\massflux_{\text{C}}=\overline{\overline{\left(\rho, u\right)}}$; the KGP form results in $\massflux_{\text{KGP}} = \overline{\rho}\,\overline{u}$. When unspecified, the KGP form is implied.}
$\checkmark$: variable preserved locally and globally,
$\times$: variable not preserved.
 \label{tab:ConvProp}}\label{tab_schemes}
\end{table}

\section{Numerical results} \label{sec:NumRes}
In this section, two and three-dimensional tests are used to study and compare the performance of the various discretizations of the compressible Euler equations obtained using different `energy' equations and forms of the convective term.

Starting with the discretization of the total energy, three classical formulations have been chosen: $(\rho E)$ is the
standard one in which the total energy equation is directly discretized with a KEP splitting of the convective term and a divergence form for the pressure term;
$(\rho E)_{\text{JP}}$ is the Jameson-Pirozzoli variant~\cite{Jameson2008b,Pirozzoli2010} in which the total enthalpy appears in the convective term by the inclusion of the scaled pressure term; $(\rho E)_{\mathrm{PEP}}$ is the PEP scheme proposed by Singh and Chandrashekar~\cite{Singh2021} and described in Section~\ref{sec:Pressure}.
In all cases the convective fluxes are discretized by using a KGP splitting.

For the internal energy equation, $(\rho e)$ is the KEP scheme using the KGP form for the convective term, which is equivalent, for exact time integration, to the KEEP schemes proposed by Kuya \emph{et al.} \cite{Kuya2018}.
The formulation $(\rho e)_{\mathrm{PEP}}$, is the PEP formulation used by Shima \emph{et al.} \cite{Shima2021}, in which the
convective term in the internal energy equation is split according to Eq.~\eqref{eq:PEP_Cond_rhoe} with $\chi=1/2$.
The formulation $(\rho e)_{\mathrm{div}}$
is the analogous PEP scheme obtained by using $\chi = 1$.
The speed of sound equation has been studied for the KEP forms F, C and KGP; they are denoted as  $(\rho c)_{\mathrm{F}}$, $(\rho c)_{\mathrm{C}}$ and $(\rho c)_{\mathrm{KGP}}$, respectively. The formulation $(\rho c)_{\mathrm{F}}$ is equivalent to the one originally proposed by Kok \cite{Kok2009}. In the same way, the three cases for the entropy equation are $(\rho s)_{\mathrm{C}}$, $(\rho s)_{\mathrm{F}}$ and $(\rho s)_{\mathrm{KGP}}$.

As regards the formulations based on the enthalpy equation,  $(\rho H)$ is the standard case in which the KGP splitting is used for the convective term. The pressure terms $\partial pu/\partial x$ and $(\gamma-1)p\partial u/\partial x$ are discretized by using standard central formulations. In addition to this case, two other scheme are analyzed in analogy with the Jameson-Pirozzoli variant for the total energy. By including the divergence part of the pressure term in the convective term for $\rho uH$ we have the form $(\rho H)_{H+\hat{p}}$; bringing the pressure term in $H$ outside of the convective term, on the other hand, results in the form $(\rho H)_{E}$.

\revone{In Table~\ref{tab:DiscretSummary} a summary of the different discretization options and of the corresponding induced terms for the `energy' equation is reported, whereas 
Table~\ref{tab_schemes} reports  the corresponding (second order) fluxes and  conservation properties.}
In all the simulations mass and momentum equations are discretized with a KEP formulation belonging to the family of forms in Eq.~\eqref{eq:Mdecomp}-\eqref{eq:Cdecomp}. The parameter $\xi$ is selected in such a way that the mass flux used in continuity and momentum  equations is the same as that in the energy equation.

The properties of the schemes have been analyzed through the study of the discrete evolution of the invariants. The $\sim$ sign over the generic variable $f$ indicates that it has been integrated over the spatial domain; the brackets denotes that it has been normalized with respect to its initial value $\widetilde{f}_0$:
\begin{equation}
    \langle \widetilde{f} \rangle = \frac{\widetilde{f} - \widetilde{f}_0}{\widetilde{f}_0}.
\end{equation}
\begin{figure}[t!]
\centering
\includegraphics[width = \linewidth]{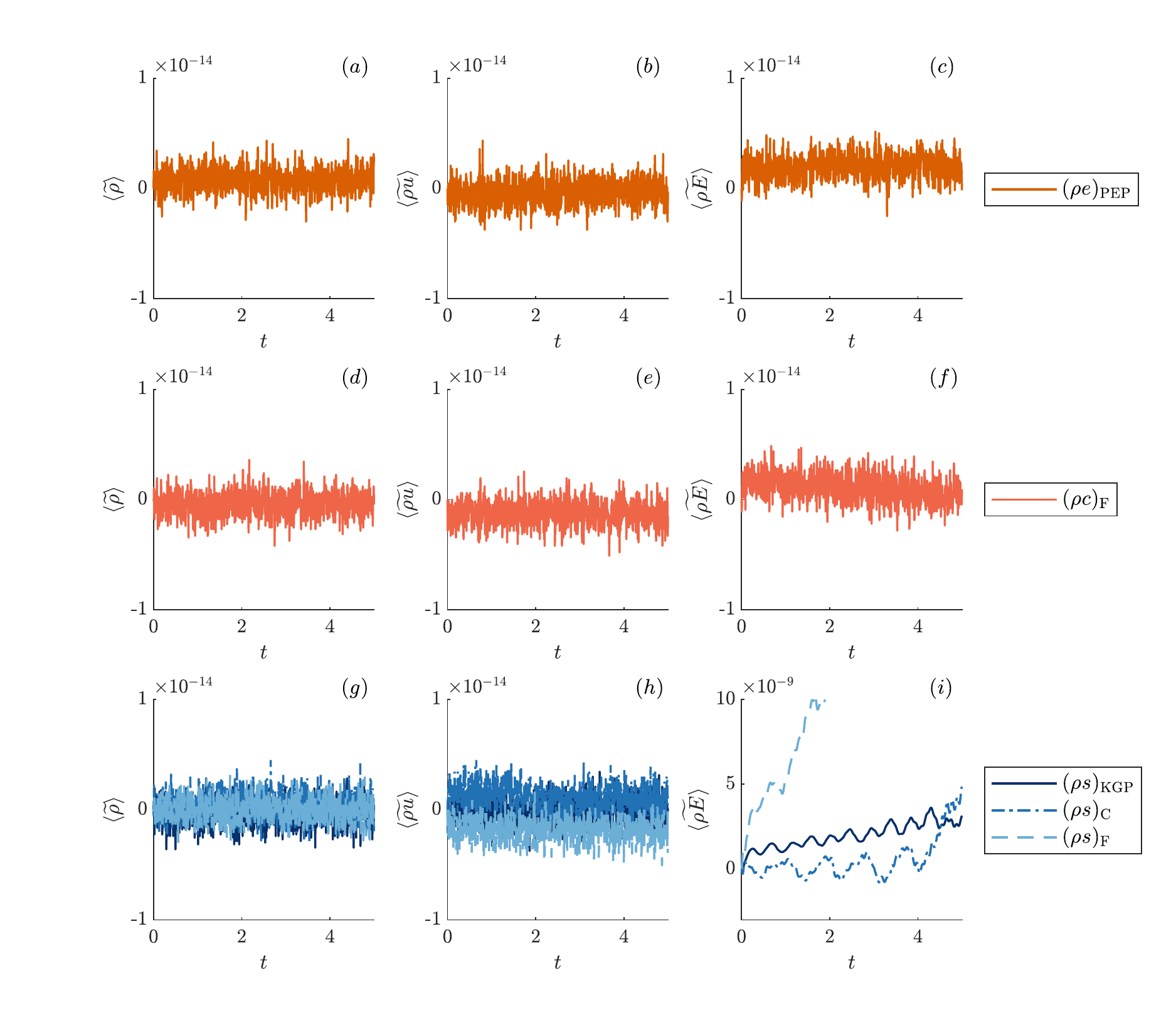}
\caption{Time evolution of linear invariants for a simulation of an isentropic vortex with a $40\times 40$ mesh using different `energy' equations: \revone{from top to bottom they are} internal energy, sound speed, and entropy equations. \revone{The invariants are, from left to right, the density, momentum and total energy integrals.} Eighth-order central schemes are employed for spatial derivatives.}\label{fig:vortex2d_comparison_linear}
\end{figure}
\subsection{Vortex advection} \label{sec:Vortex}

\begin{figure}[ht!]
\centering
\includegraphics[width = \linewidth]{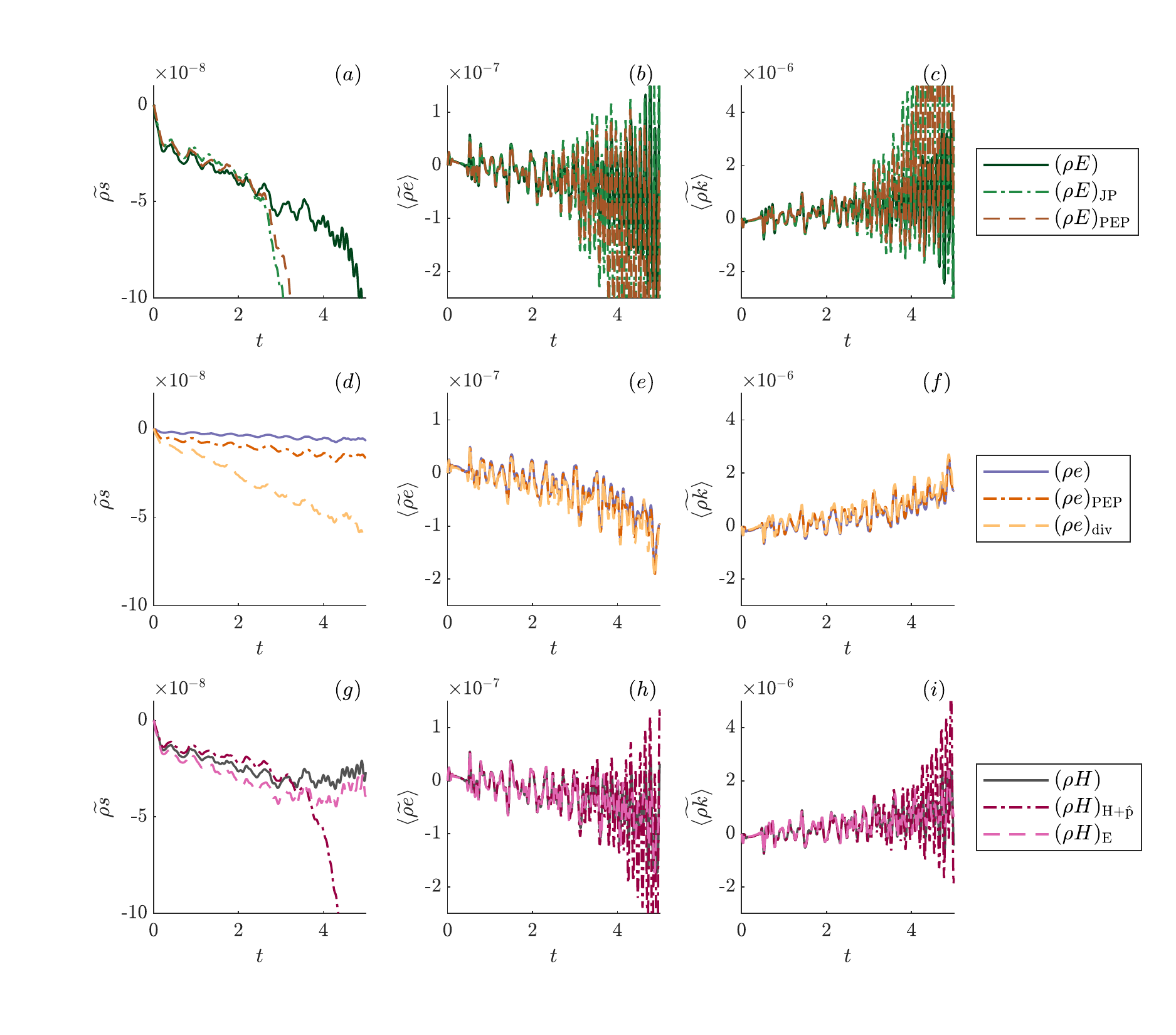}
\caption{Time evolution of integral quantities for a simulation of an isentropic vortex with a $40\times 40$ mesh using different discretizations of the total energy, internal energy, and enthalpy equations\revone{, which are shown in this order from top to bottom. From left to right, it is depicted to evolution of entropy, internal energy and kinetic energy integrals}. Eight-order central schemes are employed for spatial derivatives.}\label{fig:vortex2d_comparison_all_1}
\end{figure}
\begin{figure}[ht!]
\centering
\includegraphics[width = \linewidth]{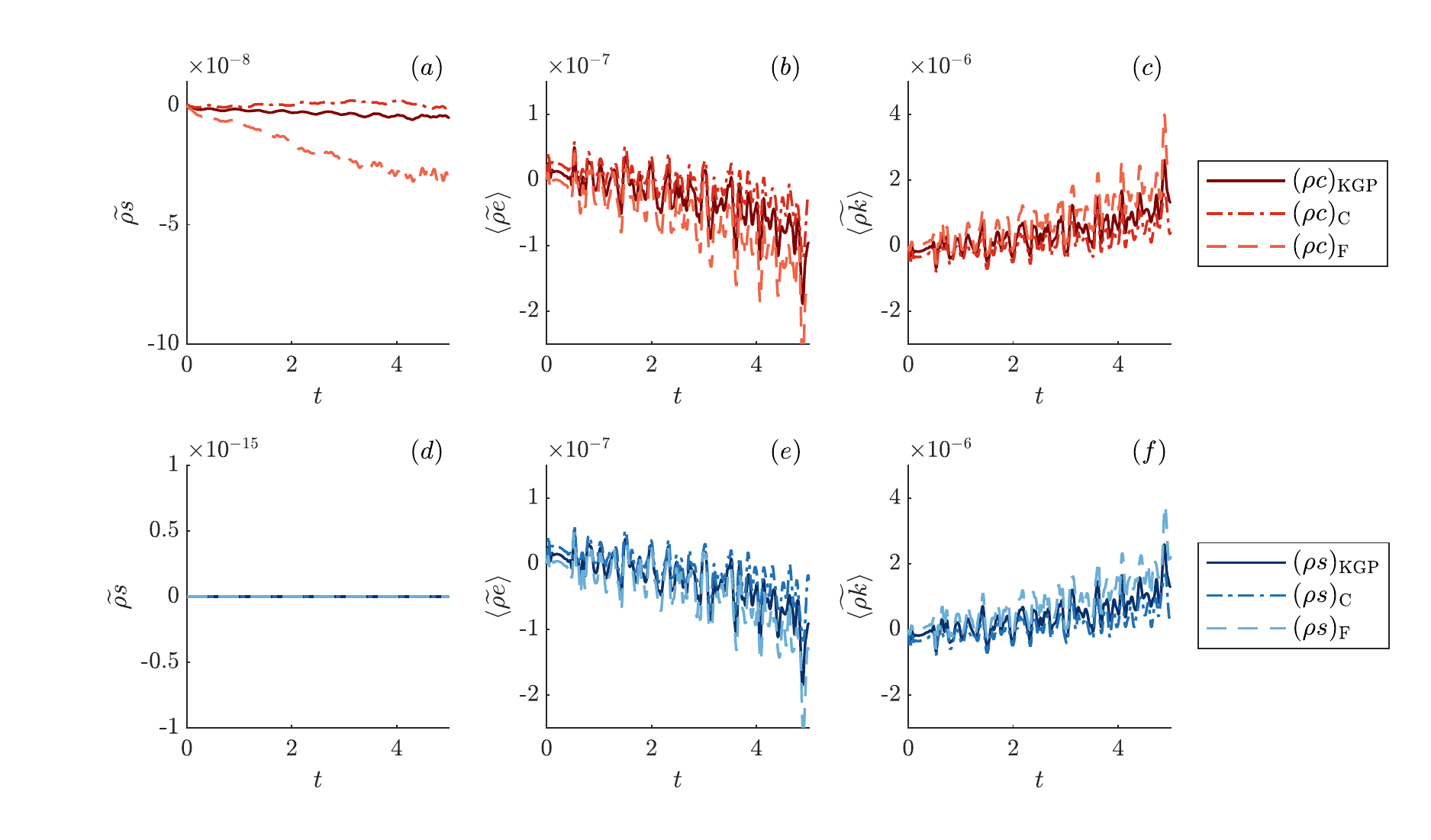}
\caption{Time evolution of integral quantities for a simulation of an isentropic vortex with a $40\times 40$ mesh using different discretizations of the sound speed and entropy equations\revone{, which are shown in this order from top to bottom. From left to right, it is depicted to evolution of entropy, internal energy and kinetic energy integrals}. Eighth-order central schemes are employed for spatial derivatives.}\label{fig:vortex2d_comparison_all_2}
\end{figure}

The two-dimensional isentropic Euler vortex problem is an exact solution of the inviscid compressible flow equations and is a commonly used test for the evaluation of the accuracy of numerical methods~\cite{Singh2021,Kok2009,SjogreenJCSC2019,Edoh2022}.
The initial conditions for the test are
\begin{align}
    \frac{ u(x,y) }{u_\infty} &= 1 - \frac{M_v}{M_\infty}\frac{y-y_0}{r_v}e^{(1-\hat{r}^2)/2}\\
    \frac{ v(x,y) }{u_\infty} &= \frac{M_v}{M_\infty}\frac{x-x_0}{r_v}e^{(1-\hat{r}^2)/2}\\
    \frac{ T(x,y) }{T_\infty} &=\left(\frac{ p(x,y) }{p_\infty}\right)^{(\gamma-1)/\gamma}  =\left(\frac{ \rho(x,y) }{\rho_\infty} \right)^{\gamma-1} = 1 - \frac{\gamma-1}{2} M_v^2 e^{1-\hat{r}^2}
\end{align}
in which $\hat{r} = r/r_v$.
A normalized value for the pressure is derived from the density as $p = \rho^\gamma / (\gamma M_\infty ^2)$.
The square domain of unitary side is discretized with a $40\times 40$ uniform Cartesian grid and boundary conditions are periodic in both directions.
The vortex, whose center has initially coordinates $(x_0,y_0) = (0.5,0.5)$, has a strength $M_v = 0.5$ and a core radius $r_v = 1/15$. The mean flow Mach number is $M_\infty = 0.5$ and the characteristic values of velocity, density, and temperature are $u_\infty =1,\ \rho_\infty=1,\   T_\infty = 720 M_v r_v$.
Time integration is performed using a standard explicit fourth-order Runge–Kutta (RK4) scheme; spatial derivatives are computed using eighth-order central schemes.
The Courant number of the tests, set to $\text{CFL} = 0.01$, corresponds to a time step size $\Delta t=1.8\times10^{-4}$.

In this test, the integral values of variables such as $\rho$, $\rho u$, $\rho E$, $\rho e$, $\rho k$, $\rho s$ should remain constant throughout the simulation, since the motion of the vortex is that of a simple translation.
The ability of the various formulations to preserve primary invariants is illustrated, for few selected cases, in Figure~\ref{fig:vortex2d_comparison_linear}.
In this and all other figures of the current section the quantities are sampled in time and displayed every $40\, \Delta t$.
In agreement with the theory, all methods were able to numerically preserve $\widetilde{\rho}$ and $\widetilde{\rho u}$ up to machine precision, since primary invariants are preserved by all the formulations used. With the only exception of the schemes discretizing the entropy equation, $\widetilde{\rho E}$ was also always preserved, as predicted by our analysis (cfr. Table~\ref{tab_schemes}). The choice of the convective term splitting for the entropy equation is of great importance, with the KGP splitting being the one that better limits the spurious total energy production in the long run (cfr. Figure~\ref{fig:vortex2d_comparison_linear}$(i)$).
\begin{figure}[ht!]
\centering
\includegraphics[width = \linewidth]{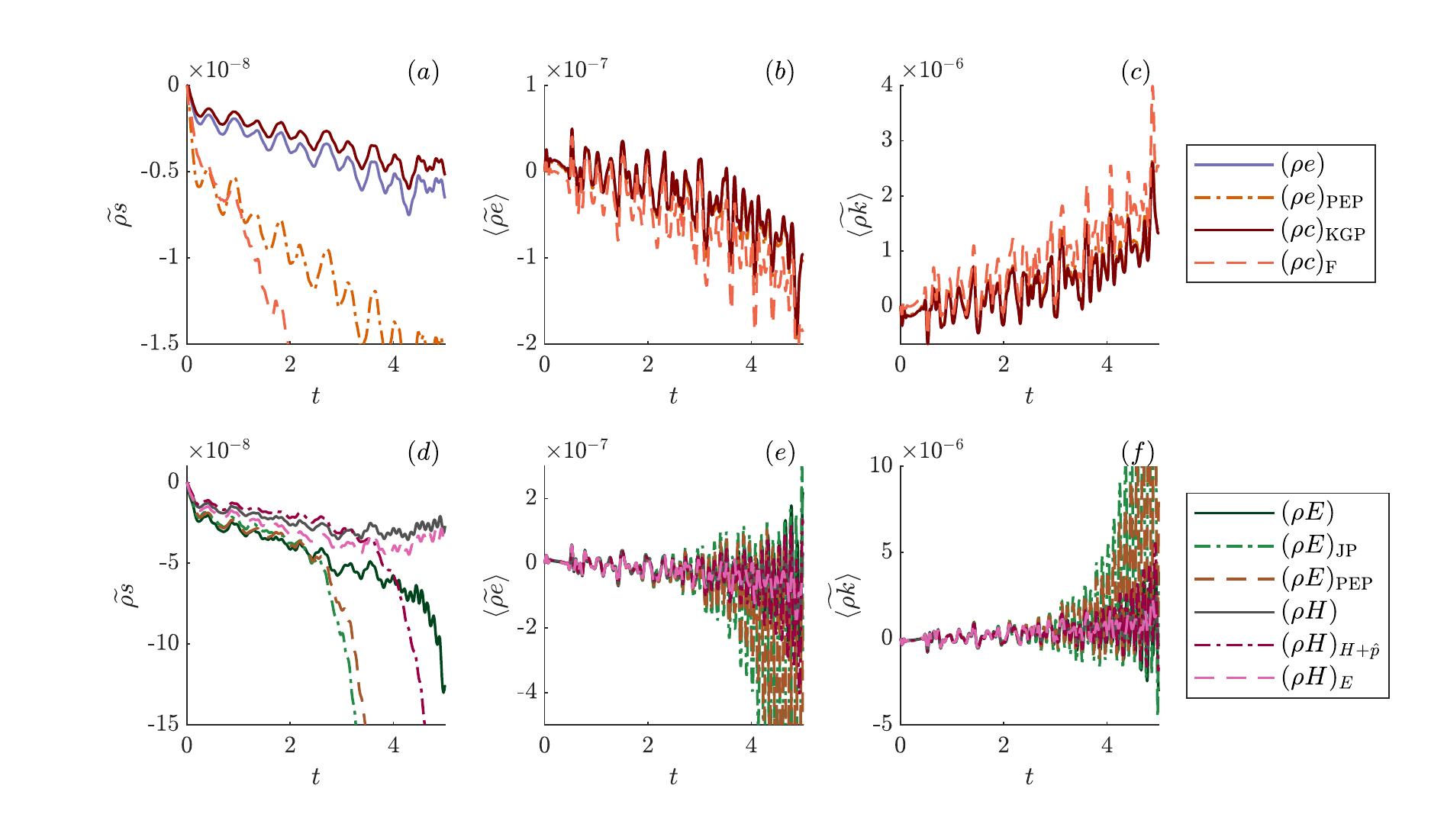}
\caption{Comparison of conservation properties of schemes based on different `energy' equations for the isentropic vortex case with a $40\times 40$ mesh and eighth-order accurate spatial discretization. On top, $(a)$-$(c)$, internal energy and sound speed schemes are compared; at the bottom, $(d)$-$(e)$, the comparison is between total energy and enthalpy schemes. \revone{From left to right, it is shown the evolution of the entropy, internal energy and kinetic energy integrals.}}\label{fig:vortex2d_comparison_C_eint_H_Etot}
\end{figure}

Considering the other invariants of the test case, a small error is present in the conservation of $\widetilde{\rho e}$ and $\widetilde{\rho k}$, with a flow between the two energies likely due to discretization errors.
The discrete evolution of the global invariants for the different methods is reported in Figures~\ref{fig:vortex2d_comparison_all_1} and \ref{fig:vortex2d_comparison_all_2}.
Due to the isentropic nature of the flow, the integral value of entropy stays constant and exactly equal to zero throughout the simulation when the entropy equation is discretized directly, irrespective of the convective term splitting.
All other methods commit an error on entropy preservation, but its magnitude greatly depends on the chosen `energy' equation and on the specific formulation.

Figure~\ref{fig:vortex2d_comparison_all_2}$(a)$ shows the improvement on entropy preservation that can be achieved by using a different convective term splitting of the sound speed equation over the F form proposed by Kok \cite{Kok2009}.
Among the formulations correctly preserving total energy, the ones using internal energy and sound speed equations showed better performances on entropy preservation. These are directly compared in Figure~\ref{fig:vortex2d_comparison_C_eint_H_Etot}$(a)$-$(c)$.

The new formulations based on the enthalpy equation display a behaviour similar to the ones based on total energy. They are compared in Figure~\ref{fig:vortex2d_comparison_C_eint_H_Etot}$(d)$-$(f)$.
\revone{From this figure}, the use of Jameson-Pirozzoli approach seems to have a negative impact (higher entropy production) in both the cases in which total energy or total enthalpy are used, in contrast to the usually reported increase of robustness for this type of formulation.
\revone{Longer simulations, however, showed that eventually the entropy error using $(\rho E)$ discretization becomes larger that the one produced by $(\rho E)_{\text{JP}}$ and $(\rho E)_{\text{PEP}}$. This happens at times $11$ and $12$ respectively. The simulation using $(\rho E)$ diverges at around time $19$, confirming the increased robustness of $(\rho E)_{\text{JP}}$ which reaches time $29$. $(\rho E)_{\text{PEP}}$ has an even longer blow-up time of $33$.}
\revone{A similar behaviour was found for the cases discretizing enthalpy, with blow-up times for $(\rho H)_{E}$, $(\rho H)$, $(\rho H)_{H + \hat{p}}$ of $22$, $26$, and $31$ respectively.}

\begin{figure}[ht]
\centering
\includegraphics[width = \linewidth]{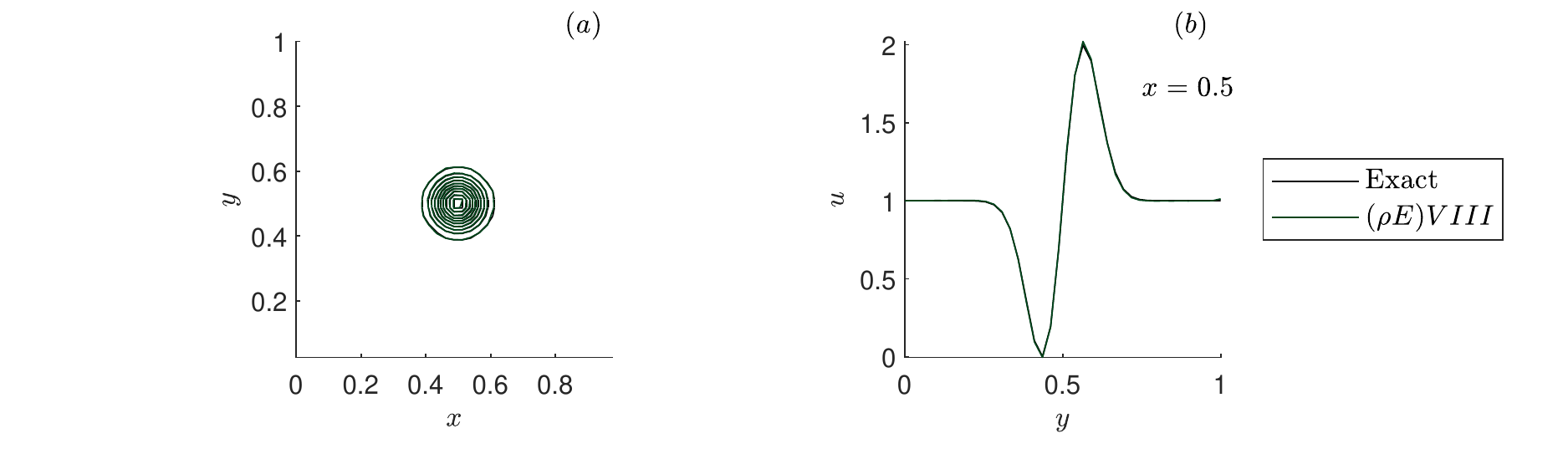} 
\includegraphics[width = \linewidth]{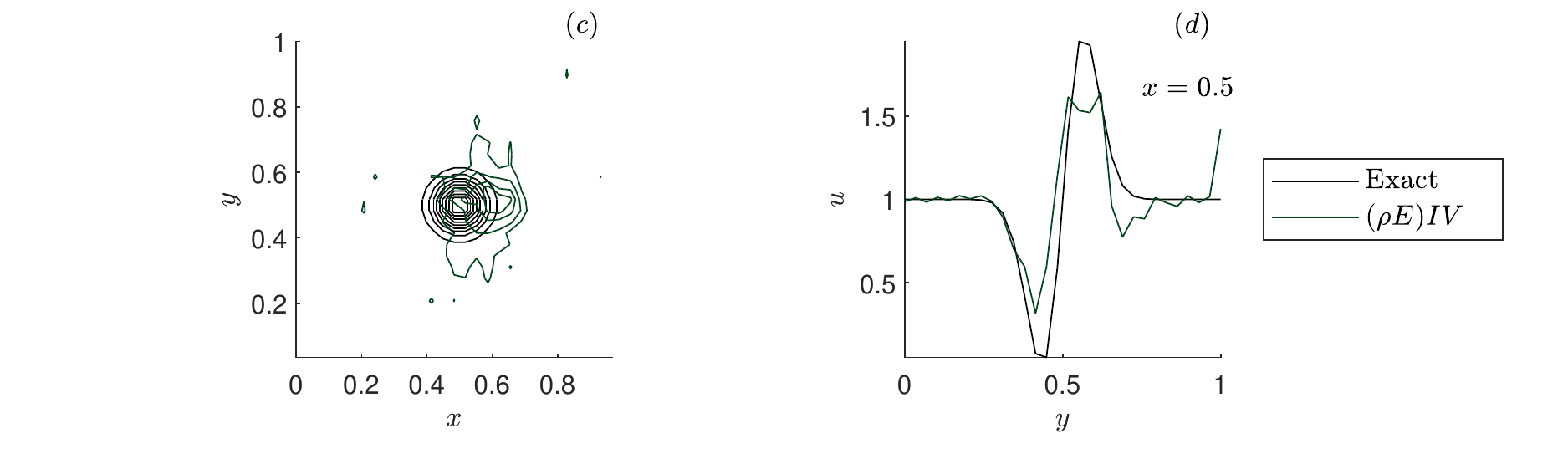} 
\caption{Comparison to the analytical solution of the isentropic vortex simulation after $T=5$. Top figures $(a)$ and $(b)$ use eighth-order discretization formulas, $\text{CFL} = 0.01$ and $40 \times 40$ mesh; bottom figures $(c)$ and $(d)$ use fourth-order discretization formulas, $\text{CFL} = 0.1$ and $30 \times 30$ mesh. On the left, $(a)$ and $(c)$ show contour lines for pressure; on the right, $(b)$ and $(d)$ represent a velocity profile}
\label{fig:vortex2d_comparison_analytical}
\end{figure}

\begin{table}[]
\centering
\begin{tabular}{@{}lccc@{}}
\toprule
                            & \multicolumn{3}{c}{\rule[0pt]{1em}{0em}Blow-up times \rule[0pt]{1em}{0em} }\\
                             & \revone{$\text{CFL} = 0.01$} & $\text{CFL} = 0.1$            & \revone{$\text{CFL} = 0.5$}     \\ \midrule
 \rule[0pt]{1em}{0em}$(\rho E)$                  & \revone{26}  & 28            & \revone{33}  \\
 \rule[0pt]{1em}{0em}$(\rho E)_{\mathrm{JP}}$    & \revone{32}      & 34           & \revone{41}  \\
 \rule[0pt]{1em}{0em}$(\rho E)_{\mathrm{PEP}}$   & \revone{46}     & 44          & \revone{--}   \\ \midrule
 \rule[0pt]{1em}{0em}$(\rho e)$                  & \revone{--}& --           & \revone{--}  \\
 \rule[0pt]{1em}{0em}$(\rho e)_{\mathrm{PEP}}$   & \revone{--}& --           & \revone{--}  \\
 \rule[0pt]{1em}{0em}$(\rho e)_{\mathrm{div}}$   & \revone{26} & 26           & \revone{34}  \\ \midrule
 \rule[0pt]{1em}{0em}$(\rho c)_{\mathrm{F}}$     & \revone{30}& 29           & \revone{35}  \\
 \rule[0pt]{1em}{0em}$(\rho c)_{\mathrm{C}}$     & \revone{20}& 19           & \revone{22}  \\
 \rule[0pt]{1em}{0em}$(\rho c)_{\mathrm{KGP}}$   & \revone{--}& --          & \revone{--}   \\ \midrule
 \rule[0pt]{1em}{0em}$(\rho H)$                  & \revone{31}& 31           & \revone{37}  \\
 \rule[0pt]{1em}{0em}$(\rho H)_{H+\hat{p}}$      & \revone{47}& 42          & \revone{--}   \\
 \rule[0pt]{1em}{0em}$(\rho H)_{E}$              & \revone{21}& 22           & \revone{28}  \\ \midrule
 \rule[0pt]{1em}{0em}$(\rho s)_{\mathrm{F}}$  & \revone{47}   & 44           & \revone{--}  \\
 \rule[0pt]{1em}{0em}$(\rho s)_{\mathrm{C}}$   & \revone{35}  & 34           & \revone{--}  \\
 \rule[0pt]{1em}{0em}$(\rho s)_{\mathrm{KGP}}$ & \revone{--}  & --           & \revone{--}  \\ \bottomrule
\end{tabular}
\caption{Blow-up times for the isentropic vortex test using different discretizations of the `energy' equation. Fourth-order accurate central schemes are employed for spatial derivatives; Courant numbers \revone{are $\text{CFL}=0.01$,} $\text{CFL}=0.1$ \revone{and $\text{CFL}=0.5$}; the Cartesian mesh is of $30 \times 30$. The symbol  `--' indicates no divergence until the end of the simulation, which is $T = 50$.}
\label{tab:blowup_time}
\end{table}
A robustness analysis was also executed \revone{for all cases} on longer simulations with an end time $T = 50$ and \revone{$\text{CFL} = 0.01,0.1$ and $0.5$}, for which fourth-order derivation schemes are employed and the domain is discretized with a $30\times 30$ mesh. As shown in Figure~\ref{fig:vortex2d_comparison_analytical}, the lower accuracy results in a degraded solution, but the schemes display similar conservation properties.
The blow-up times for the various formulations are reported in Table~\ref{tab:blowup_time}. The use of Jameson-Pirozzoli approach and the PEP formulation resulted in an incresed robustness; however no formulation using total energy or enthalpy equation was able to reach the end time of the simulation, \revone{except for the formulations $(\rho E)_{\text{PEP}}$ and $(\rho H)_{H + \hat{p}}$ at the highest CFL number}.
\revone{Higher robustness within the integration time at all CFL} was only achieved by the schemes using the KGP form in the discretization of the  entropy, sound speed and internal energy equations and by the PEP formulation of internal energy proposed by Shima \emph{et al.}~\cite{Shima2021}.
\revone{Comparing the results at $\text{CFL} = 0.1$ with those at $0.5$, it is evident a general increase in robustness when a larger time step is used. This is likely due to the fact that the employed time integrator introduces a slight dissipative error.
When considering $\text{CFL} = 0.01$, this result is not as clear. This can be attributed to the fact that, for this value, the error due to the temporal integrator is already very small and other effects may have a larger impact on robustness.
The effect of the temporal error, however, does not seem to change the relative robustness of the discretization choices when comparing them to each other.}

\revtwo{Considering the blow-up times at $\text{CFL} = 0.01$ for the total energy ($(\rho E)$: $19$, $(\rho E)_{\text{JP}}$: $29$, $(\rho E)_{\text{PEP}}$: $33$) and enthalpy($\,(\rho H)_{E}$: $22$, $(\rho H)$: $26$, $(\rho H)_{H + \hat{p}}$: $31$) obtained using $40\times 40$ mesh and eighth-order accurate derivation schemes it is also possible to draw some conclusions about the impact of the resolution. Blow-up times are lower when compared at the same $\text{CFL}$ number with the simulations using $30\times30$ mesh and fourth-order accurate derivation schemes (Table~\ref{tab:blowup_time}). This suggests that an increased resolution can lower the blow-up time of the simulations.}

\subsection{3D Taylor-Green vortex} \label{sec:Taylor}
The inviscid Taylor-Green vortex is a classical benchmark widely used for the evaluation of discretization schemes for turbulence simulations, as it includes the generation of small scales by three-dimensional vortex stretching and transition to randomized flow, while allowing the evaluation of the spurious entropy production due to the numerical scheme.
The domain consists in a triperiodic box of side $2\pi$, which is discretized into a grid of $32\times32\times32$ nodes.
The initial condition is given by
\begin{align}
    \rho(x,y,z) &= 1\\
    u(x,y,z) &= \sin(x)\cos(y)\cos(z)\\
    v(x,y,z) &= -\cos(x)\sin(y)\cos(z)\\
    w(x,y,z) &= 0\\
    p(x,y,z) &= 100 + \frac{(\cos(2x) + \cos(2y))(\cos(2x)+2)-2}{16}
\end{align}
in which the pressure value is chosen to be sufficiently high to keep the flow nearly incompressible, with a Mach number $M<0.1$.
The spatial derivatives are discretized through explicit fourth-order accurate central schemes; the temporal integrator is the standard RK4. The Courant number of the simulations, based on the initial conditions, is $\text{CFL} = 0.1$.
Contrary to the implementation used for the two-dimensional test, all schemes used the total energy as the primary variable, but with the appropriate induced equation to emulate the different cases.
\revtwo{For exact time integration the two approaches are equivalent. In real simulations some differences could be triggered by temporal errors, which in our tests are minimized by the small CFL number.}
\begin{figure}[ht!]
\centering
\includegraphics[width = \linewidth]{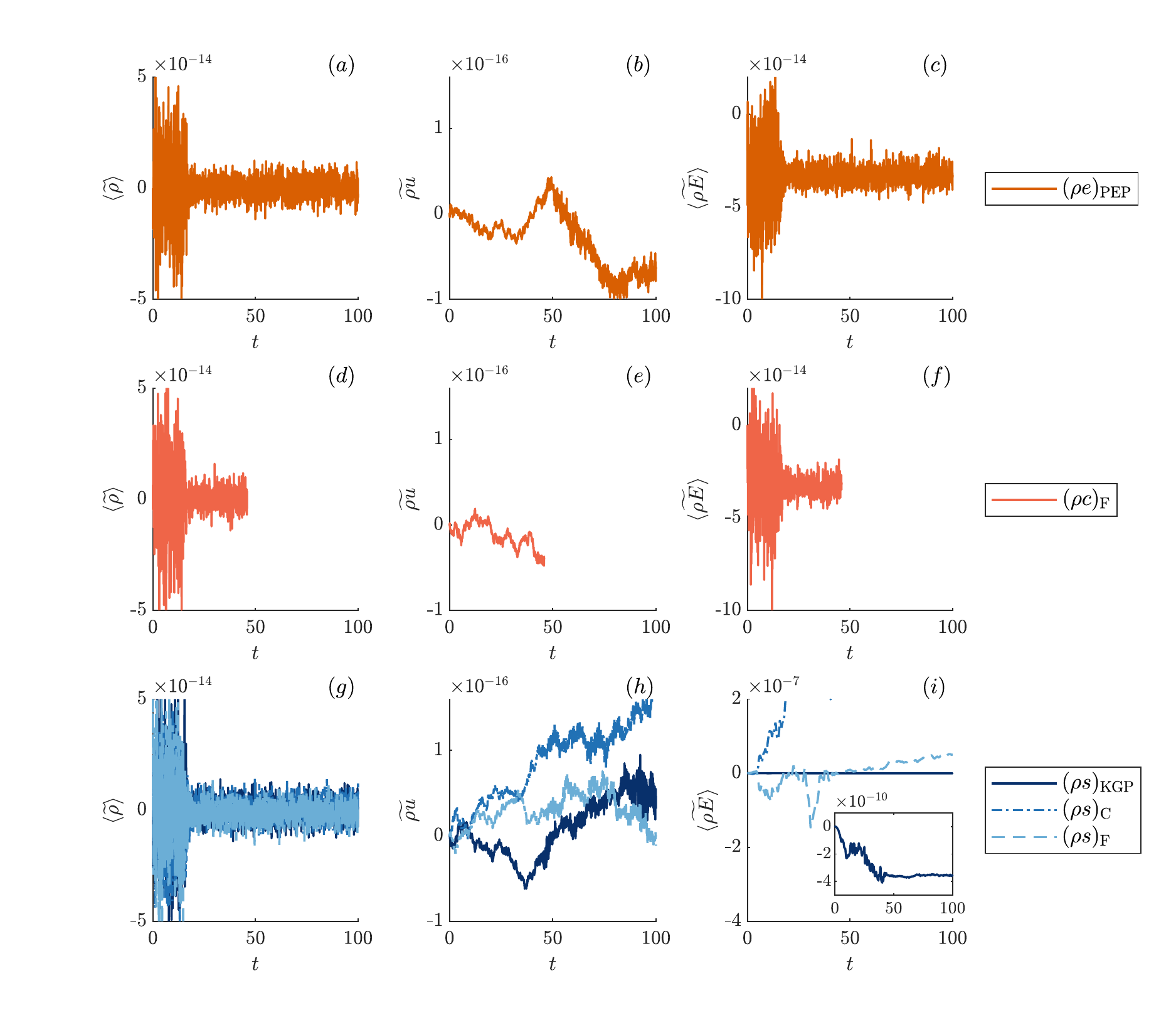}
\caption{Time evolution of linear invariants for the inviscid Taylor-Green vortex with a $32\times 32\times 32$ mesh using different `energy' equations: \revone{from top to bottom they are} internal energy, sound speed, and entropy equations. \revone{The invariants are, from left to right, the density, momentum and total energy integrals.}
\revtwo{Subfigure~$(i)$ shows in greater details the behaviour of the case $(\rho s)_{\text{KGP}}$ in its inset.} 
Fourth-order central schemes are employed for spatial derivation.}
\label{fig:TGV3d_comparison_linear}
\end{figure}

For the inviscid Taylor-Green flow the integral values of $\rho$, $\rho u$, $\rho E$, and  $\rho s$ are expected to stay constant; on the other hand the integral of $\rho e$, $\rho k$ can change since their evolution is affected by energy exchanges.
The evolution of the primary invariants $\rho,\rho u$ and $\rho E$ for selected formulations is reported in Figure~\ref{fig:TGV3d_comparison_linear}.
To display the data of this test, the quantities have been sampled every $10\, \Delta t$.
As in the previous test, all formulations conserved the values of $\widetilde{\rho}$ and $\widetilde{\rho u}$. This was also the case for $\widetilde{\rho E}$ for all the methods that did not discretize the entropy equation directly.
Note that in Figure~\ref{fig:TGV3d_comparison_linear}$(d)$-$(f)$ the plots are truncated at $t\simeq 50$, due to the blow up of the simulation for the $(\rho c)_{\text{F}}$ formulation.
The plots show that the numerical preservation of linear invariants is retained up to the blow-up time.
From Figure~\ref{fig:TGV3d_comparison_linear}$(i)$ it is possible to appreciate the comparatively better energy preserving property of $(\rho s)_{\text{KGP}}$, with the error on the normalized integral value at time 100 being of the order of $10^{-10}$, while it reaches $5\times10^{-8}$ for $(\rho s)_{\text{F}}$ and $3\times10^{-4}$ for $(\rho s)_{\text{C}}$.

\begin{figure}[ht]
\centering
\includegraphics[width = \linewidth]{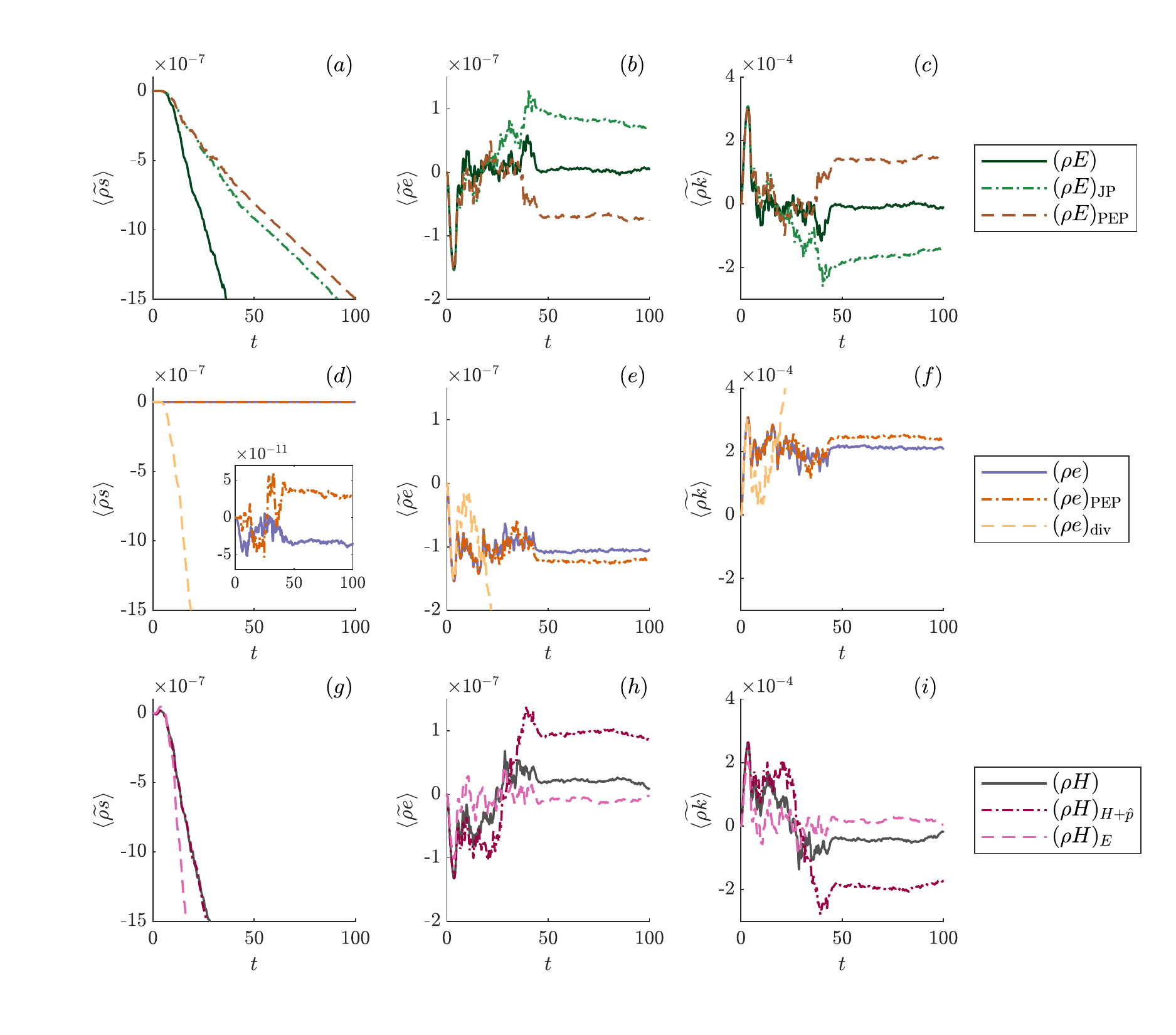} 
\caption{Time evolution of integral quantities for the inviscid Taylor-Green vortex with a $32\times 32\times 32$ mesh using different discretizations of the total energy, internal energy and enthalpy equations\revone{, which are shown in this order from top to bottom. From left to right, it is depicted to evolution of entropy, internal energy and kinetic energy integrals}. \revtwo{Subfigure~$(d)$ shows in greater details the behaviour of the cases $(\rho e)$ and $(\rho e)_{\text{PEP}}$ in its inset.} Fourth-order central schemes are employed for spatial derivation.}
\label{fig:TGV3d_comparison_all_1}
\end{figure}
\begin{figure}[ht]
\centering
\includegraphics[width = \linewidth]{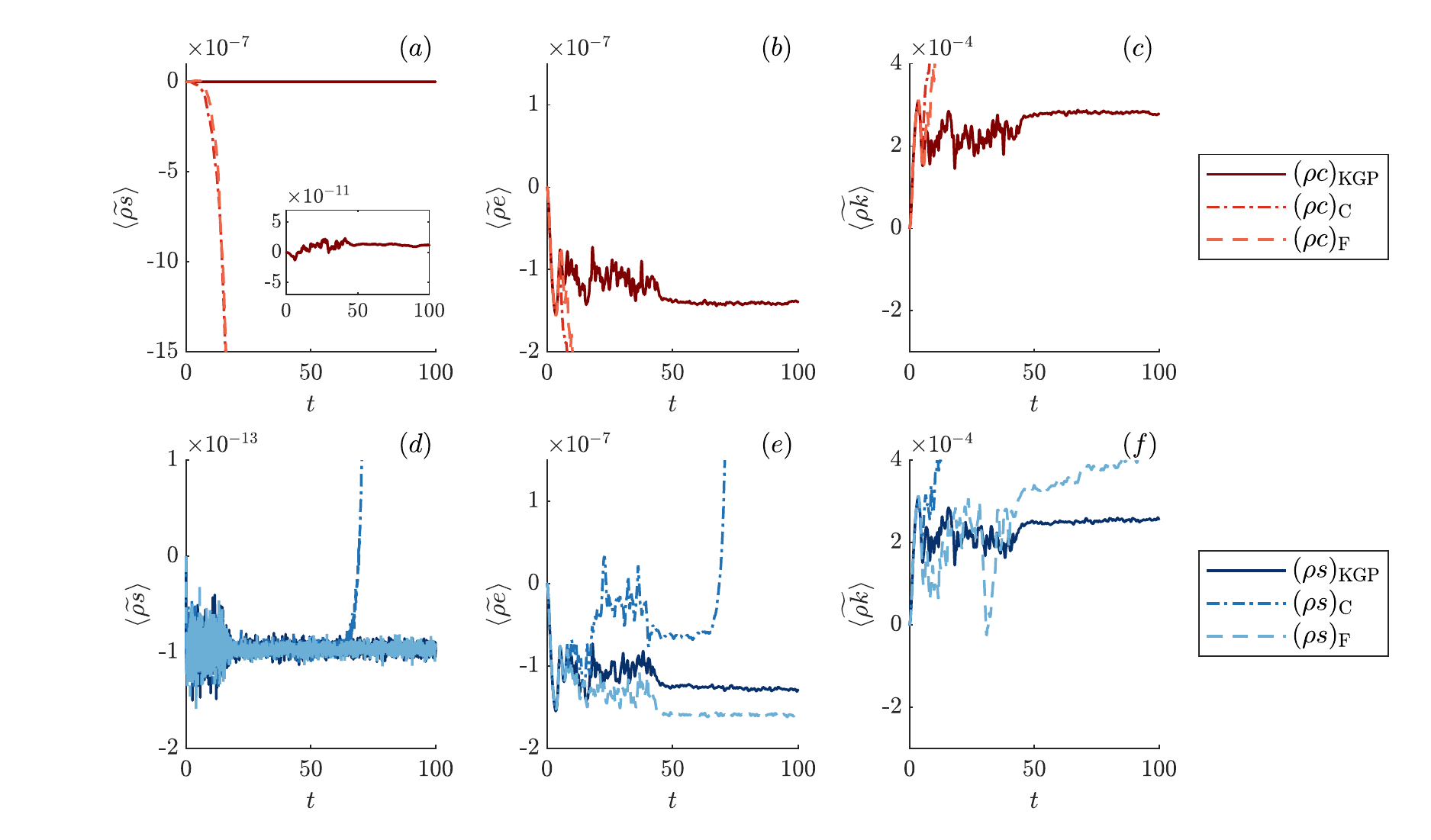} 
\caption{Time evolution of integral quantities for the inviscid Taylor-Green vortex with a $32\times 32\times 32$ mesh using different discretizations of the sound speed and entropy equations\revone{, which are shown in this order from top to bottom. From left to right, it is depicted to evolution of entropy, internal energy and kinetic energy integrals}. \revtwo{Subfigure~$(a)$ shows in greater details the behaviour of the case $(\rho c)_{\text{KGP}}$ in its inset.} Fourth-order central schemes are employed for spatial derivation.}
\label{fig:TGV3d_comparison_all_2}
\end{figure}

Figures~\ref{fig:TGV3d_comparison_all_1} and \ref{fig:TGV3d_comparison_all_2} show the discrete evolution of other global quantities.
The performance of the various formulation on entropy preservation is consistent with what was seen from the previous test.
In Figure~\ref{fig:TGV3d_comparison_all_2}$(d)$ the scheme $(\rho s)_{\text{C}}$ exhibits a deviation from the expected exact
conservation of entropy. As already observed in
\cite{Coppola2019}, this behaviour has to be attributed
to the use of the emulated total energy equation in place of the entropy equation, which causes the deterioration of the performances
of the scheme (which, however, remains stable within the simulation time used)  affecting also the preservation of the linear invariant $\widetilde{\rho s}$.

\begin{figure}[ht]
\centering
\includegraphics[width = \linewidth]{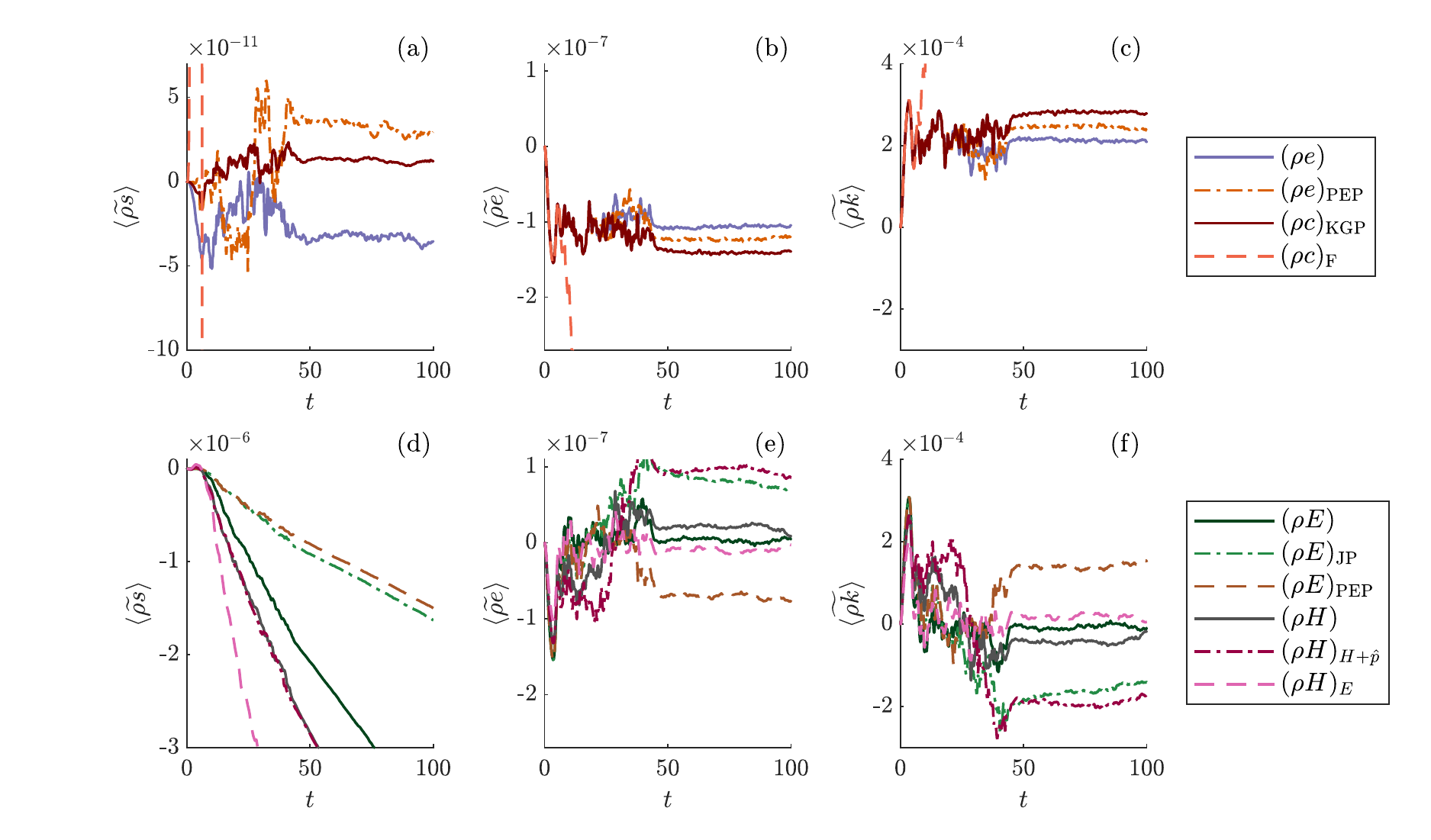}
\caption{Comparison of conservation properties of schemes based on different `energy' equations for the Taylor-Green vortex case with a $32\times 32\times 32$ mesh and fourth-order accurate spatial discretization. On top, $(a)$-$(c)$, internal energy and sound speed schemes are compared; at the bottom, $(d)$-$(e)$, the comparison is between total energy and enthalpy schemes. \revone{From left to right, it is shown the evolution of the entropy, internal energy and kinetic energy integrals.}}
\label{fig:TGV3d_comparison_C_eint_H_Etot}
\end{figure}

Among the other formulations, the best entropy preservation performances were shown by schemes using internal energy and sound speed equations; a comparison is shown in Figure~\ref{fig:TGV3d_comparison_C_eint_H_Etot}.
For this test, in addition to $(\rho e)$ and $(\rho c)_{\text{KGP}}$, the PEP formulation $(\rho e)_{\text{PEP}}$ also managed to have a small spurious entropy production. The $(\rho c)_{\text{C}}$ and $(\rho c)_{\text{F}}$ were the only two formulations to diverge, with blow-up times of 45 and 46 respectively.

Figure~\ref{fig:TGV3d_comparison_C_eint_H_Etot} also compares total energy and entropy formulations. In contrast with what was seen for the isentropic vortex case the Jameson-Pirozzoli approach has \revone{from the beginning} a positive effect on the entropy production when using the total energy equation; on the other hand it did non change the behaviour when using the enthalpy equation. The PEP formulation showed some benefits as well when used on the total energy equation.

\begin{figure}[ht]
\centering
\includegraphics[width = \linewidth]{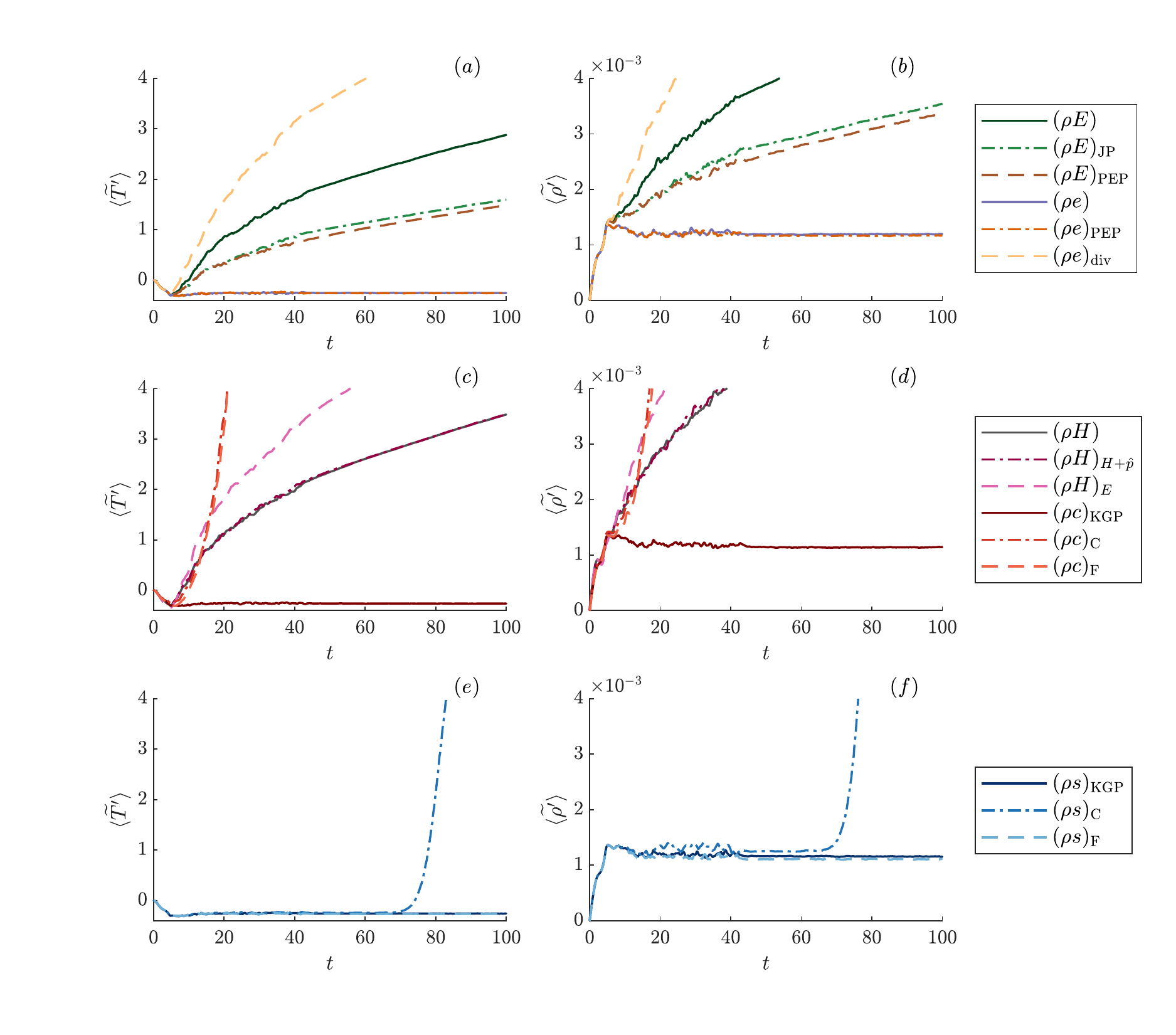}
\caption{Temperature \revone{(on the left)} and density \revone{(on the right)} fluctuations for the inviscid Taylor-Green vortex using different `energy' equation formulations. \revone{On top, $(a)$-$(b)$, total and internal energy schemes are compared; in the middle, $(c)$-$(d)$, the comparison is between enthalpy and sound speed schemes;  at the bottom, $(e)$-$(f)$, it is shown the behaviour for the enthalpy schemes.} The mesh size is of $32 \times 32 \times 32$ and and fourth-order central schemes are employed for spatial derivation.}
\label{fig:TGV3d_comparison_flut}
\end{figure}

With this test case it is also possible to analyze the time evolution of thermodynamic fluctuations and gain an insight into the reliability of the simulations. After an initial transient, these are expected to stabilize around a constant value, as in the case of inviscid isotropic homogeneous turbulence~\cite{HoneinMoin2004,Pirozzoli2010,Coppola2019}.
The evolution of temperature and density fluctuations, $ \langle \rho '\rangle $ $ \langle T '\rangle $, is displayed in Figure~\ref{fig:TGV3d_comparison_flut}.
An increase in the fluctuations with time is exhibited by all formulations based on total energy or enthalpy equations. The sound speed equation is capable of reaching an asymptotically constant level of fluctuations when used with a KGP splitting and the same result was found for the internal energy. This was also achieved by the PEP formulation $(\rho e)_{\text{PEP}}$. When using the entropy equation the C form failed to contain the increase in the fluctuations, while both the F and KGP forms had them stabilized around a constant value.


\revtwo{Additional tests on the TGV case for a higher value of the
initial Mach number ($M\simeq0.3)$ were also carried out. Up to around $t\simeq 70$ the simulations basically confirm the findings of the $M\simeq0.1$ case, with a slightly increase of the exchanges of energy
between kinetic and internal energies, due to the pressure-work terms.
For $t\geq70$ almost all the simulations show a deviation from the lower Mach number case, basically characterized by a (potentially unbounded) increase of the fluctuations, as
it occurs for the formulation $(\rho s)_C$ in Figure~\ref{fig:TGV3d_comparison_flut}$(e)$-$(f)$.
This picture is in accordance with the findings of Honein and Moin~(\cite{HoneinMoin2004}, p.~542), who for a similar test case report the same behavior and attribute it to the
eventual occurrence of shocks, even after several turnover times.
Low resolution and the absence of shock capturing schemes cause the build up of instabilities and a divergence of the fluctuations from the equilibrium value. }

 \section{Conclusions}\label{sec:Conclusions}
The discrete conservation properties of various formulations of the compressible Euler equations have been analyzed, both theoretically and numerically.
The analysis has been carried out within a semi-discretized approach and has focused on the numerical treatment of the energy equation within a finite-difference (or an equivalent finite-volume) framework.
Two important factors have been considered: the energy variable whose equation is directly discretized and the split form used for  convective and pressure derivatives.
The theoretical analysis has been  conducted by studying the discrete evolution equations for selected thermodynamic quantities, as they are induced by the adopted discretization for primary  variables. The relations among the discrete convective and pressure terms in the various equations have been investigated to infer general criteria under which additional induced conservation properties can be obtained.
Some of the most popular formulations used in the literature have been considered and some new ones have been also proposed.
A detailed  analysis on the locally-conservative character of the induced kinetic-energy equation shows that, except for the case in which the entropy equation is directly discretized, all the  analyzed methods can be reformulated as an equivalent locally-conservative formulation involving the discretization of the total energy. In all these cases entropy is not strictly preserved and a discrete evolution equation for entropy has been derived.

Numerical tests have been conducted on two classical benchmarks widely used in the literature and all the theoretical predictions have been confirmed. Moreover, some additional robust behaviours have been detected, giving an impression of the general performances of the various methods with respect to induced conservation properties.
Among the various options considered, the formulations involving a direct discretization of the internal energy or entropy equations show the most robust and accurate results in the numerical tests.
In the first case both KEP formulations (preserving $\rho e^2$) or square-root preserving formulations (preserving $\rho \sqrt{e}\propto \rho c$) have better conservation and robustness properties when used in conjunction to the KGP splitting, as compared to analogous discretizations involving total energy or enthalpy. The internal energy formulation has shown better properties than the total energy analogues also when PEP schemes are used.

In the case in which the entropy equation is directly discretized, numerical tests show that the adoption of the KGP splitting enhances the favourable properties of the discretization, confirming previous results from the literature.
However, formulations based on locally-conservative entropy discretizations do not induce in general a locally-conservative discretization of the total energy equation.

\revone{The final choice that has to be made among the various alternatives presented in this paper could depend on additional factors that have not been considered in this work. Some of them could be of practical character, as the simple improvement of an existing code written by using a certain set of primitive variables. Other could be more conceptual, as the modeling of the SGS terms stemming from the filtered equations in the various approaches in a LES framework.
The choice of which induced quantity is more useful to consider and the correct way to numerically preserve in the various applications is hence still a topic which deserves further investigation.
}

The proposed analysis suggests also several further developments, which could constitute future topics of investigation.
The usually adopted strategy of employing a KEP discretization for the energy equation could be not justified, since the preservation of other derived quantities (i.e. the square root) confers similar robustness to the formulation, allowing the reproduction of more physical mechanisms at a discrete level. \revone{Moreover, the proposed analysis indicates also the possibility of building KEP procedures by using point-dependent formulations, with a consequent increase in the number of degrees of freedom,
which can be optimized to achieve different targets.}

\appendix
\section{}\label{Sec:Appendix}
The high-order extension of the theory exposed in Sec.~\ref{sec:LocalConservation} and \ref{sec:Discrete_KE_Eq} can be
illustrated by considering the full expression for the fluxes in Eq.~\eqref{eq:Flux3}.
By substituting the relations \eqref{eq:InterpOp} into Eq.~\eqref{eq:Flux3} one has for the convective term in the mass equation the expression
\begin{equation}
    \mM = \xi\left\llbracket 2\sum_{k=1}^La_k\sum_{m=0}^{k-1}\overline{\rho u}^\revtwo{i-m+k/2}\right\rrbracket+
    (1-\xi) \left\llbracket 2\sum_{k=1}^La_k\sum_{m=0}^{k-1}\overline{\overline{\left(\rho,u\right)}}^\revtwo{i-m+k/2}\right\rrbracket
\end{equation}
from which the mass flux can be written as
\beq
\mF_{\rho} =  2\sum_{k=1}^La_k\sum_{m=0}^{k-1}\revtwo{m_{i-m+k/2}}
\eeq
where $\revtwo{m_{i+k/2}} = \xi\overline{\rho u}^\revtwo{\,i+k/2} + (1-\xi)\overline{\overline{\left(\rho, u\right)}}^\revtwo{\,i+k/2}$.
The convective flux for $\rho\phi$ can be obtained in the same way by substituting Eq.~\eqref{eq:InterpOp_Cforms}
into Eq.~\eqref{eq:Flux3}
\beq\label{eq:HighOrdFlux}
\mF_{\rho\phi} =  2\sum_{k=1}^La_k\sum_{m=0}^{k-1}\overline{\phi}^\revtwo{{\,i-m+k/2}}\revtwo{m_{i-m+k/2}}.
\eeq
The higher-order generalized kinetic energy flux, analogous to Eq.~\eqref{eq:KinEn_Flux}, can be
calculated starting from the expression for the convective term of the generalized kinetic energy
in Eq.~\eqref{eq:ConvKinGlobPres}. According to the definitions of the interpolation operators
in Eq.~\eqref{eq:InterpOp} the associated numerical flux is given by
\beq
\mF_{\rho\phi^2/2} =  2\sum_{k=1}^La_k\sum_{m=0}^{k-1}\left[\left(\frac{\xi}{2}\right)\overline{\overline{\left(\phi,\rho u\phi\right)}}^\revtwo{{\,i-m+k/2}} + \left(\frac{1-\xi}{2}\right)\overline{\overline{\left(u\phi,\rho \phi\right)}}^\revtwo{{\,i-m+k/2}}\right],
\eeq
which can be manipulated as follows:
\begin{multline}
    \mF_{\rho\phi^2/2} =  2\sum_{k=1}^La_k\sum_{m=0}^{k-1}\left[\frac{\xi}{2}
    \left(\frac{\phi_{i-m}\left(\rho u\phi\right)_{i-m+k} + \phi_{i-m+k}\left(\rho u\phi\right)_{i-m}}{2}\right)\right.+ \\
    \left.\frac{1-\xi}{2}\left(\frac{\left(u\phi\right)_{i-m}\left(\rho\phi\right)_{i-m+k} + \left(u\phi\right)_{i-m+k}\left(\rho \phi\right)_{i-m}}{2}\right)\right]=\\
    2\sum_{k=1}^La_k\sum_{m=0}^{k-1}\left[\frac{\phi_{i-m}\phi_{i-m+k}}{2}\left(\xi\overline{\rho u}^\revtwo{{\,i-m+k/2}} + (1-\xi)\overline{\overline{\left(\rho, u\right)}}^\revtwo{{\,i-m+k/2}}\right)\right].
\end{multline}
The final form of the higher-order kinetic energy flux can be expressed as
\beq\label{eq:HighOrdFlux_KinEn}
\mF_{\rho\phi^2/2} =  2\sum_{k=1}^La_k\sum_{m=0}^{k-1}\frac{\phi_{i-m}\phi_{i-m+k}}{2}\revtwo{\,m_{i-m+k/2}}.
\eeq
Note that Eq.~\eqref{eq:HighOrdFlux} and \eqref{eq:HighOrdFlux_KinEn} have been used also in \cite{Kuya2021} for the
case in which the mass flux is discretized with the skew-symmetric form obtained with $\xi=1/2$.

\FloatBarrier

\bibliographystyle{elsarticle-num-names}
\bibliography{biblio}
\end{document}